\documentclass[journal=nalefd,manuscript=letter,layout=twocolumn]{achemso}
\usepackage[version=3]{mhchem} % Formula subscripts using \ce{}
\usepackage{graphicx}% Include figure files
\usepackage{dcolumn}% Align table columns on decimal point
\usepackage{bm}% bold math
\usepackage[mathlines]{lineno}% Enable numbering of text and display math
\usepackage{amssymb}
\usepackage{amsmath}% bold math
\usepackage{soul,color}
\title{Can CF$_3$-functionalized La@C$_{60}$
be isolated experimentally and
become superconducting?}

\author{Jie~Guan}
\affiliation{Physics and Astronomy Department,
             Michigan State University,
             East Lansing, Michigan 48824, USA}

\author{David Tom\'{a}nek}
\affiliation{Physics and Astronomy Department,
             Michigan State University,
             East Lansing, Michigan 48824, USA}
\email%[E-mail: ]%
            {tomanek@pa.msu.edu}%

\date{\today} % delete this line to display the current date

\abbreviations{DFT, PBE, PAW, HOMO, LUMO, HPLC}

\keywords{metallofullerene, La@C$_{60}$, $\it{ab~initio}$,
electronic structure
\\}

\begin{document}

%%%%%%%%%%%%%%%%%%%%%%%%%%%%%%%%%%%%%%%%%%%%%%%%%%%%%%%%%%%%%%%%%%%%%
%% The manuscript does not need to include \maketitle, which is
%% executed automatically.  The document should begin with an
%% abstract, if appropriate.  If one is given and should not be, the
%% contents will be gobbled.
%%%%%%%%%%%%%%%%%%%%%%%%%%%%%%%%%%%%%%%%%%%%%%%%%%%%%%%%%%%%%%%%%%%%%

\begin{abstract}
Superconducting behavior even under harsh ambient conditions is
expected to occur in La@C$_{60}$ %
%\hl{%
if it could be isolated from the primary metallofullerene soot
when functionalized by CF$_3$ radicals. %
%}%
We use {\em ab initio} density functional theory calculations to
compare the stability and electronic structure of C$_{60}$ and the
La@C$_{60}$ endohedral metallofullerene to their counterparts
functionalized by CF$_3$. We found that CF$_3$ radicals favor
binding to C$_{60}$ and La@C$_{60}$, and have identified the most
stable isomers. Structures with an even number $m$ of radicals are
energetically preferred for C$_{60}$ and structures with odd $m$
for La@C$_{60}$ due to the extra charge on the fullerene. This is
consistent with a wide HOMO-LUMO gap in La@C$_{60}$(CF$_3$)$_m$
with odd $m$, causing extra stabilization in the closed-shell
electronic configuration. CF$_3$ radicals are both stabilizing
agents and molecular separators in a metallic crystal, which could
increase the critical temperature for superconductivity.
\end{abstract}

%%%%%%%%%%%%%%%%%%%%%%%%%%%%%%%%%%%%%%%%%%%%%%%%%%%%%%%%%%%%%%%%%%%%

%\section*{Introduction}

The discovery of the C$_{60}$ fullerene~\cite{C60Kroto} with a
hollow cage structure immediately triggered the question, whether
the space inside could be filled by other atoms. This question has
been answered affirmatively shortly afterwards by successfully
encapsulating metal atoms including Li, Na, K, Rb, Sr, Ba, Sc, La,
Tc and U inside M@C$_{2n}$ endohedral fullerenes, also called
metallofullerenes.~\cite{%
{Heath1985},% La, K, Ba and Sr
{Chai1991},% La and K
{Haufler90},% U -- this has 17 authors
{Shinohara1993jpc},% Sc
{Tellgmann96},% Li, Na, K and Rb
{CAMPBELL19971763},% Li, Na, K and Rb
{Almeida96},% N claimed endohedral
{KNAPP1997433},% N claimed endohedral
{Mauser97},% N claimed endohedral
{Hirata96},% K
{Weck-TcC60}, %Tc
{Shinohara2000}% review
}%

A major reason for the current interest in metallofullerenes such
as La@C$_{60}$ is the possibility to observe superconductivity in
solid C$_{60}$ that is doped endohedrally rather than exohedrally
by 3 electrons~\cite{Shinohara-private}. A molecular crystal
formed of isolated and recrystallized La@C$_{60}$ molecules would
be electronically related to M$_3$C$_{60}$ with M representing
alkali atoms. In M$_3$C$_{60}$ crystals, superconductivity with
$T_c{\lesssim}40$~K has been observed and explained by
electron-phonon coupling that is modulated by the lattice
constant~\cite{{DT053},{DT057}}. The same behavior is expected to
occur in the isoelectronic La@C$_{60}$ crystal. Unlike
alkali-based C$_{60}$ superconductors, La@C$_{60}$ crystals will
be stable under ambient or even harsh conditions.

In spite of the fact that fullerene cages of different size favor
encapsulation of metal atoms energetically~\cite%
{{Chai1991},{Shinohara1993jpc},%
{Shinohara2000},{Yamamoto1994},{KIKUCHI199367}}, most of the
resulting complexes are highly reactive due to their open-shell
configuration and form insoluble polymerized
solids~\cite{wang2013}. Only M@C$_{2n}$ metallofullerenes with
large cages~\cite{Shinohara1993jpc} ($2n{\gtrsim}74$) have been
successfully isolated from the raw soot, with M@C$_{82}$
dominating. It has been difficult to extract the most abundant
M@C$_{60}$ and M@C$_{70}$ metallofullerenes due to their
insolubility in regular fullerene
solvents~\cite{{Shinohara2000},{Ogawa2000},{Kubozono96}} such as
toluene and CS$_2$. Even though several metallofullerenes with
cages as small as C$_{60}$ have been extracted by solvents
such as pyridine and aniline~\cite{{WANG1993354},{Wang1993},%
{Kubozono1995},{Kubozono96}}, it has become a large challenge to
separate the M@C$_{60}$ fraction, since pyridine and aniline
solvents are not suitable for high-performance liquid
chromatography (HPLC)~\cite{Ogawa2000}. Only recently, La@C$_{70}$
and Y@C$_{2n}$ with $2n{\geq}60$, functionalized by CF$_3$
radicals, have been separated using toluene and CS$_2$ as
solvents~\cite{{wang2013},{wang2016}}. In this new strategy, the
function of trifluoromethyl radicals was to further separate the
metallofullerene cages and to chemically stabilize their unstable
open-shell electronic configuration~\cite{Diener1998}. Isolation
and crystallization of the more interesting La@C$_{60}$ molecules
is currently being attempted~\cite{Shinohara-private}.

% Executive summary

In view of the possibility to obtain superconducting behavior in
crystalline La@C$_{60}$ even under harsh ambient conditions, we
provide theoretical support for the experimental effort to
solubilize the La@C$_{60}$ endohedral metallofullerene from the
primary soot. We made use of {\em ab initio} density functional
theory calculations to compare the stability and electronic
structure of the bare C$_{60}$ fullerene and La@C$_{60}$ to their
counterparts functionalized by $m$ CF$_3$ radicals. We found that
several CF$_3$ radicals can form stable bonds to C$_{60}$ and
La@C$_{60}$, and have identified the most stable structural
isomers for $m{\leq}5$. Generally, structures with an even $m$ are
energetically preferred for C$_{60}$ and structures with an odd
$m$ for La@C$_{60}$ due to the extra 3 electrons donated by the
encapsulated La. This is consistent with our finding that a wide
HOMO-LUMO gap opens in La@C$_{60}$(CF$_3$)$_m$ molecules with an
odd $m$ value, causing extra stabilization in the closed-shell
electronic configuration. The two-fold function of CF$_3$ radicals
as stabilizing agents and molecular separators may help in
isolating specific molecules from the primary soot. If
La@C$_{60}$(CF$_3$)$_m$ could be solubilized and recrystallized,
endohedral instead of exohedral doping by three electrons per
C$_{60}$ and use of stabilizing CF$_3$ radicals as spacers may
lead to superconductivity with a relatively high $T_c$ achieved by
%\hl{%
increasing
%}%
the lattice constant of the molecular
crystal~\cite{{DT053},{DT057}}. %
%\hl{%
We should note that the range of favorable lattice constants is
limited, since their increase beyond a critical value changes
doped C$_{60}$ from a metal to a Mott-Hubbard %
insulator{~\cite{{Ganin-10},{Nomura-eC60-15}}}.
%}%
Even more appealing %
%\hl{%
than changing the lattice constant
%}%
appears the possibility to remove CF$_3$ radicals after obtaining
La@C$_{60}$(CF$_3$)$_m$ in solution and
subsequently recrystallize La@C$_{60}$. %
Stability even under harsh ambient conditions is expected due to
the absence of alkali atoms filling
interstitial sites in known M$_3$C$_{60}$ superconductors. %
Moreover, the possibility to encapsulate different elemental
species provides an additional handle on tuning
$T_c$ of M@C$_{60}$ superconductors.{~\cite{Takeda-ArC60}} %

%===========< FIGURE 1 >=========================================
\begin{figure}[t]
\includegraphics[width=1.0\columnwidth]{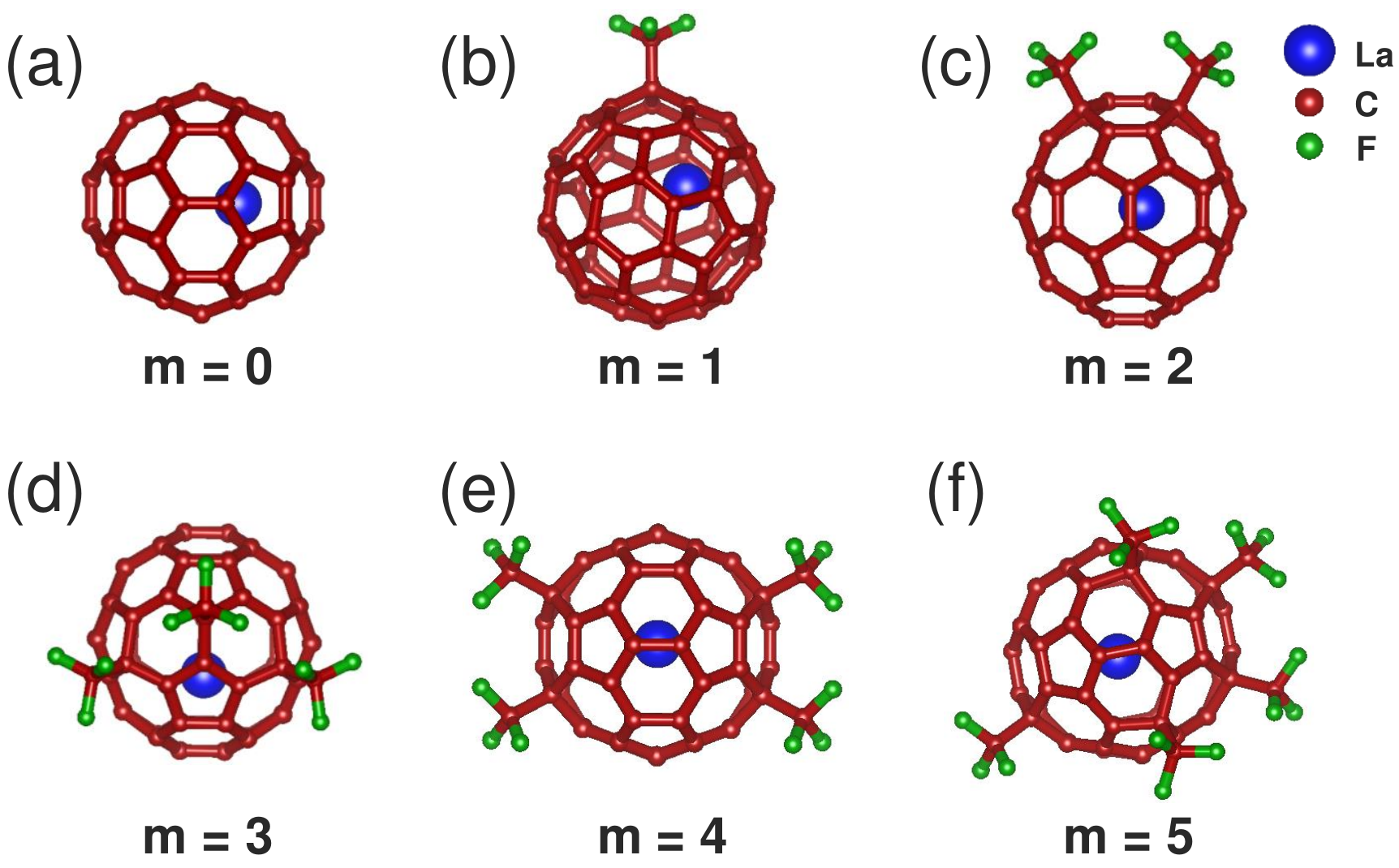}
\caption{(Color online) Ball-and-stick models of the optimized
structure of the most stable isomers of La@C$_{60}$(CF$_3$)$_m$
molecules with (a) $m=0$, (b) $m=1$, (c) $m=2$, (d) $m=3$, (e)
$m=4$, (f) $m=5$. \label{fig1} }
\end{figure}
%===========< FIGURE 1 >=========================================

%===========< FIGURE 2 >=========================================
\begin{figure}[t]
\includegraphics[width=0.9\columnwidth]{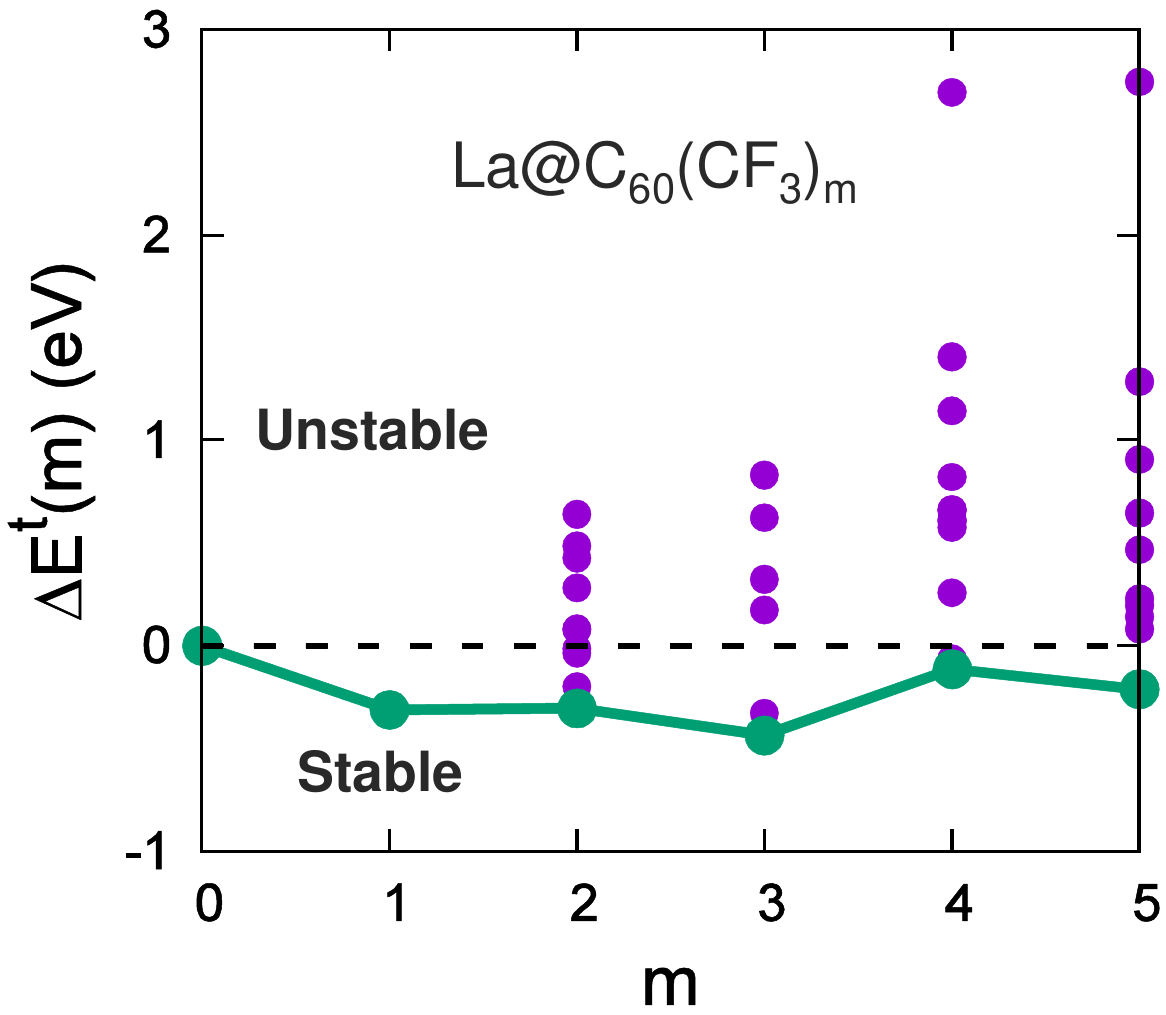}
\caption{(Color online) Energy change ${\Delta}E^t(m)$
% based on DFT-PBE
associated with attaching an extra trifluoromethyl radical to the
most stable La@C$_{60}$(CF$_3$)$_{m-1}$ isomer.
% Defined by
% ${\Delta}E^t(m)=E_{tot}(m)%
% -E_{tot}^{min}(m-1)%
% -(1/2)E_{tot}[$(CF$_3$)$_2]$. %
% $E_{tot}^{min}(m)$ denotes the total energy of the most stable
% La@C$_{60}$(CF$_3$)$_m$ isomer
Negative values denote stable structures and ${\Delta}E^t(0)$ is
set to zero. ${\Delta}E^t(m)$ values for the most stable isomers
are shown by the larger symbols and are connected by %
the solid green line to guide the eye.%
\label{fig2} }
\end{figure}
%===========< FIGURE 2 >=========================================

\section{Results}%

% Equilibrium geometry for most stable isomers

We have used density functional theory (DFT) with the
Perdew-Burke-Ernzerhof (PBE)~\cite{PBE} exchange-correlation
functional to determine the equilibrium geometry of La@C$_{60}$
and La@C$_{60}$(CF$_3$)$_m$ molecules for $m{\leq}5$. For a given
number $m$ of attached CF$_3$ radicals, we optimized many
different regioisomers, as specified in the Supporting
Information, and displayed the most stable structures in
Figure~\ref{fig1}.
% For $m{\geq}2$, the trifluoromethyls can have more than one
% configuration on the C$_{60}$ surface. Many different isomers for
% $m$ from 2 to 5 are calculated and the detailed results are
% discussed in Appendix A. Here we show the most stable isomers for
% $m$ from 2 to 5 in Figure~\ref{fig1}(c)-(f).
Figure~\ref{fig1}(a) displays the bare C$_{60}$ molecule
containing a La atom in its equilibrium off-center configuration.
As seen in Figure~\ref{fig1}(b), CF$_3$ radicals prefer to attach
on-top of C atoms in the C$_{60}$ cage. Whereas the adsorption of
a single CF$_3$ radical on the bare C$_{60}$ cage in a metastable
geometry is an endothermic process requiring $0.43$~eV, presence
of La in the La@C$_{60}$ metallofullerene turns this adsorption
process exothermic with an energy gain of $0.31$~eV. We considered
10 regioisomers for $m=2$ and display their optimum geometry, the
corresponding Schlegel diagram and relative energy in Figure~S1
of the Supporting Information, %
with the most stable geometry given in Figure~\ref{fig1}(c). These
results indicate that CF$_3$ radicals in the most stable $m=2$
regioisomer are in para (third neighbor) positions on a single
hexagon on the C$_{60}$ surface, and that this structure has a
C$_{2v}$ symmetry. Other arrangements of the radicals are
penalized energetically up to ${\lesssim}1$~eV. Comparing the
relative energies of other regioisomers, we found that CF$_3$
radicals prefer to be close, but not too close on the C$_{60}$
surface.

A very similar picture emerges for $m=3$ CF$_3$ radicals adsorbed
on La@C$_{60}$. Six different regioisomers, their relative
energies and the corresponding Schlegel diagrams are presented in
Figure~S2 %
of the Supporting Information. As for $m=2$, the optimum
arrangement of CF$_3$ radicals in the most stable isomer, shown in
Figure~\ref{fig1}(d), is in para position on adjacent hexagonal
rings on the C$_{60}$ surface and results in a mirror symmetry for
the molecule. The second most stable isomer contains CF$_3$
radicals separated by 5 neighbor distances, is only
${\approx}0.1$~eV less stable than the most stable structure, and
has a C$_3$ symmetry. Even though nearest-neighbor arrangements of
CF$_3$ radicals were not considered, other structural candidates
incurred an energy penalty of up to ${\lesssim}1.3$~eV with
respect to the most stable isomer.

The structural paradigm changes when increasing the number of
CF$_3$ radicals attached to La@C$_{60}$ to $m=4$. Among the 10
regioisomers presented in Figure~S3 %
of the Supporting Information, %
the most stable structure, shown in Figure~\ref{fig1}(e), contains
two pairs of CF$_3$ radicals in para-arrangement on hexagonal
rings that are separated by half the circumference of the C$_{60}$
molecule. Another arrangement, with all CF$_3$ radicals in para
arrangement on adjacent hexagonal rings, is energetically the
second-best isomer, with its energy only $0.051$~eV higher than
the most stable structure. The most stable isomer has a C$_{2v}$
symmetry and the second-best isomer only a mirror symmetry.

The structural paradigm for La@C$_{60}$(CF$_3$)$_m$ regioisomers
with $m=5$ CF$_3$ radicals is similar to the $m=4$ case. The most
stable of ten regioisomers, presented in Figure~S4 %
of the Supporting Information and displayed in
Figure~\ref{fig1}(f), contains 4 CF$_3$ radicals in para
arrangement on three adjacent hexagonal rings. The last radical is
separated by 4 neighbor distances from the closest CF$_3$ radical.
It is also possible to arrange all five CF$_3$ radicals in para
positions on four adjacent hexagonal rings, but this highly
symmetric regioisomer is less stable by ${\lesssim}0.9$~eV than
the most stable structure.

% Preferential number of CF$_3$s

To find out the preferential number of CF$_3$ radicals attached to
the La@C$_{60}$ molecule, we calculated the energy change
${\Delta}E^t(m)$ associated with attaching an extra
trifluoromethyl radical to the most stable
La@C$_{60}$(CF$_3$)$_{m-1}$ isomer and display our results in
Figure~\ref{fig2}. We defined ${\Delta}E^t(m)$ by
\begin{equation}
{\Delta}E^t(m)=
E_{tot}(m)%
-E_{tot}^{min}(m-1)%
-\frac{1}{2}E_{tot}({\rm C}_2{\rm F}_6) \,. %
\label{eq1}
\end{equation}
$E_{tot}^{min}(m)$ denotes the total energy of the most stable
La@C$_{60}$(CF$_3$)$_m$ isomer. A negative value of
${\Delta}E^t(m)$ indicates an exothermic reaction for adsorbing an
extra CF$_3$ radical that had been initially formed by
dissociating a hexafluoroethane molecule. ${\Delta}E^t(0)$ for the
bare metallofullerene has been set to zero.

According to our results in Figure~\ref{fig2}, it is always
possible to find a structural arrangement of CF$_3$ radicals that
would further stabilize the La@C$_{60}$ structure. We find
structures with odd number of radicals to be relatively more
stable, with $m=3$ providing the optimum stabilization with
${\Delta}E^t(3)=-0.44$~eV. In comparison, the stability gain
${\Delta}E^t(4)=-0.11$~eV upon adsorbing four radicals is much
smaller. Equally interesting as the optimum regioisomer is the
range of ${\Delta}E^t(m)$ values for a given $m$, which is as wide
as $3$~eV. We should not neglect the fact that functionalization
by CF$_3$ radicals within the raw soot may not always yield the
most stable isomers.

%===========< FIGURE 3 >=========================================
\begin{figure*}
\centering
\includegraphics[width=1.6\columnwidth]{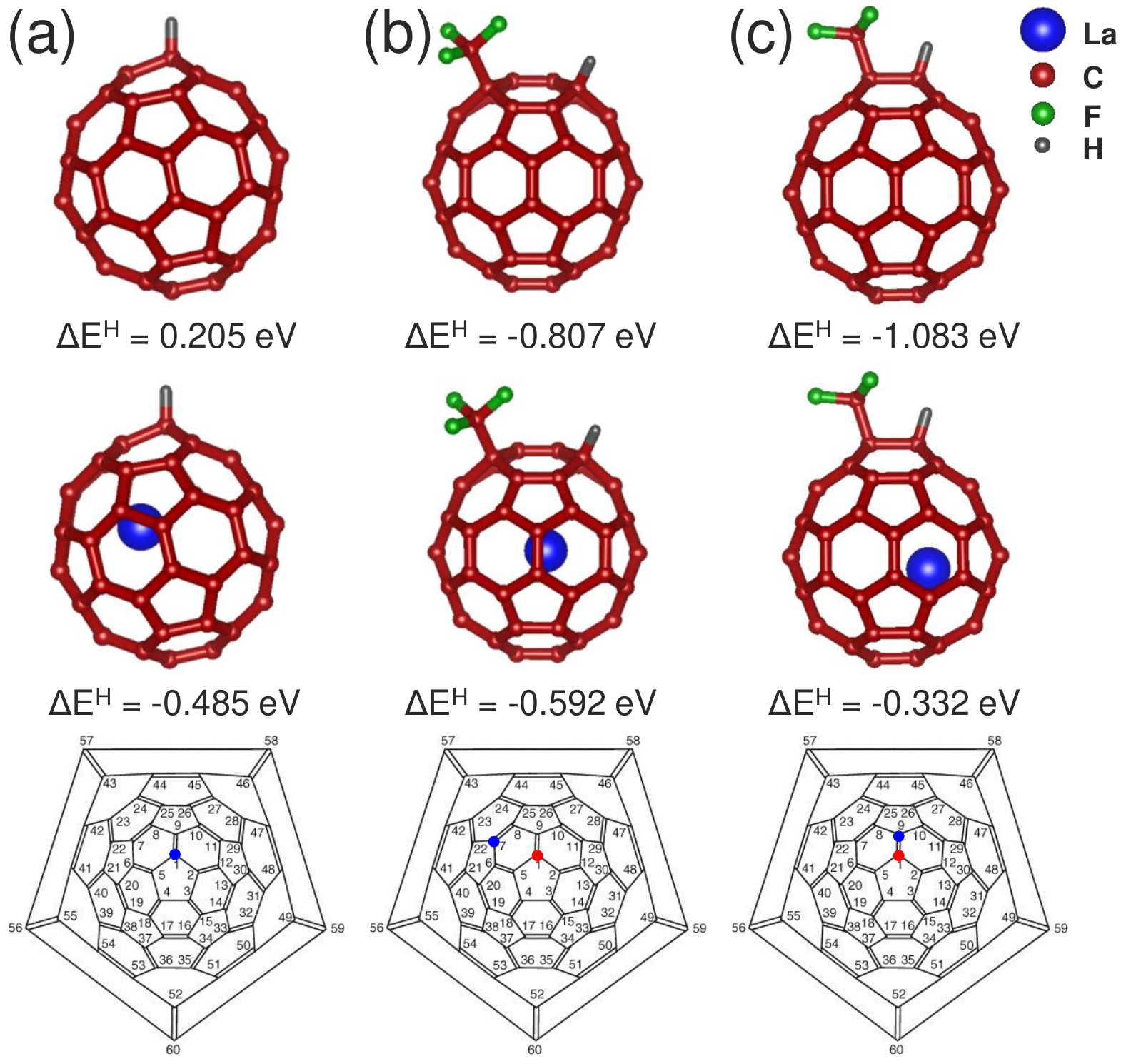}
\caption{(Color online) Structural and energetic change caused by
hydrogen atom adsorption on (a) a C$_{60}$ (top panel) and
La@C$_{60}$ (middle panel), and (b,c) on C$_{60}$CF$_3$ (top
panels) and La@C$_{60}$CF$_3$ (middle panels). The CF$_3$ radical
and the H atom are adsorbed on the same hexagonal ring, occupying
(b) para (or third-neighbor) positions or (c) ortho (or
first-neighbor) positions. The bottom panels contain the
corresponding Schlegel diagrams of C$_{60}$ with the hydrogen
sites indicated by the blue and the CF$_3$ sites by the red dots.
${\Delta}E^H$ denotes the energy change due to hydrogenation, with
H$_2$ as the hydrogen source and negative values denoting an
exothermic process.
% Hydrogenated La@C$_{60}$(CF$_3$)$_m$ for (a) $m=0$, $m=1$ with
% hydrogen and trifluoromethyl at (b) para-position and (c)
% ortho-position in a hexagon. The first row is the ball-and-stick
% model of the structures without La atom inside the cage and the
% second row is the structures with La atom inside the cage. The
% last row is the corresponding Schlegel diagram of C$_{60}$ with
% hydrogen site indicated by blue dot and trifluoromethyl site by
% red. ${\Delta}E^H$ here is defined as ${\Delta}E^H=
% E_{tot}$(La@C$_{60}$(CF$_3$)$_m$-H)$ -
% E_{tot}$(La@C$_{60}$(CF$_3$)$_m$)$ -1/2E_{tot}$(H$_2$), or
% ${\Delta}E^H=E_{tot}$(C$_{60}$(CF$_3$)$_m$-H)
% $-E_{tot}$(C$_{60}$(CF$_3$)$_m$)$-1/2E_{tot}$(H$_2$) for that
% without La atom inside.
\label{fig3} }
\end{figure*}
%===========< FIGURE 3 >=========================================

The even-odd alternation in the ${\Delta}E^t(m)$ values for the
most stable isomers, with odd values referring to higher
stability, has been found previously for CF$_3$-functionalized
La@C$_{70}$ and Y@C$_{70}$ molecules~\cite{wang2013,wang2016}. The
even-odd alternation is also found in CF$_3$-functionalized
fullerenes, which, however, display an energetic preference for an
even number of CF$_3$ radicals~\cite{Darwish2004,trifluo2007}. The
preference of fullerenes for an even number of CF$_3$ radicals can
be explained easily. Attaching an even number of radicals to
nominal double-bonds in the fullerene cage converts the C atoms in
the double-bond from an $sp^2$ to an $sp^3$ configuration, leaving
no radical behind. Consequently, an odd number of CF$_3$ radicals
will leave at least one C radical behind, lowering the stability
of the structure.

As we will discuss next, an La atom -- upon its encapsulation --
transfers 3 electrons to the C$_{60}$ shell, whereas the charge
transfer caused by attached CF$_3$ radicals is significantly
smaller. The odd number of electrons added to the shell modifies
its electronic structure, resulting in an energetic preference for
an odd instead of an even number $m$ of CF$_3$ radicals attached
to C$_{60}$.
% This difference is caused by the encapsuled metal atom La, which
% has three valence electrons. The odd number of extra free
% electrons will try to transfer to the cage and modify the
% electronic environment of the surface, which consequently results
% in a preference for an odd number of attached CF$_3$s.

Electronically, there is no difference if the negative charge of
the C$_{60}^{3-}$ shell stems from encapsulated La in a
hypothetical La@C$_{60}$ or from 3 alkali atoms that are
interstitial in the M$_3$C$_{60}$ crystal. The electronic
structure of the corresponding molecular crystal will be
determined by the narrow band formed of the partly occupied LUMO
of the C$_{60}$ molecules. Except for a dynamical Jahn-Teller
effect, the system should be metallic and superconducting, with
$T_c$ depending in the same way on the lattice constant as in the
equivalent M$_3$C$_{60}$ molecular crystal~\cite{DT057}. %
The present understanding of the origin of superconductivity in
alkali-based M$_3$C$_{60}$ solids is discussed in the Supporting
Information.%

Plausibilizing the difference between the functionalization of a
bare C$_{60}$ fullerene and the La@C$_{60}$ metallofullerene
requires understanding the charge redistribution introduced by the
encapsulated La atom in C$_{60}$. %
We discuss this charge redistribution in detail in the Supporting
Information, in particular in Figure~S1. The essence of our
findings is consistent with the %
Bader charge analysis~\cite{Bader06,Bader07,Bader09,Bader11} of
the bare C$_{60}$ molecule and all metallofullerenes depicted in
Figure~\ref{fig1}. Independent of the presence and number $m$ of
attached CF$_3$ radicals, we found a net transfer of
${\approx}1.8$ electrons from the encapsulated La atom to the
C$_{60}$ cage. Restating this finding in a different way, the net
charge of the La atom was not affected by the presence of CF$_3$
radicals carrying a net Bader charge $Q{\approx}-0.1e$, which was
transferred locally from the C$_{60}$ cage alone. In view of the
ambiguity to assign a delocalized charge to particular atoms, we
may assume that the charge transfer from the La atom to the cage
is likely underestimated by the Bader analysis. Comparing to
similar systems, where a Bader analysis has been
performed~\cite{{Lee2009prb},{DT249}}, we found the enclosed La to
be most likely in the $3+$ oxidation state, with the C$_{60}$ cage
carrying extra 3 electrons.

% Bader charge analysis for CF$_3$ radicals and La in
% La@C$_{60}$(CF$_3$)$_m$
% m=0:                                    Q(La) = +1.81 e
% m=1: Q(CF$_3$) = -0.14 e (extra electrons) Q(La) = +1.79 e
% m=2: Q(CF$_3$) = -0.12 e - isomer 2(1)  Q(La) = +1.79 e
% m=5: Q(CF$_3$) = -0.10 e - isomer 5(2)  Q(La) = +1.77 e
% m=5: Q(CF$_3$) = -0.13 e - isomer 5(8)  Q(La) = +1.77 e

The net electron transfer from the encapsulated atom to the
surrounding cage is also the cause of the off-center displacement
of La by ${\approx}1.2$~{\AA}. In DFT-PBE, this symmetry-lowering
displacement is associated with an energy gain of
${\approx}3.3$~eV. This stabilization energy can be simply
explained by the electrostatic polarization energy gain caused by
a point charge moving off-center inside a spherical metallic
shell, as discussed previously for other
metallofullerenes~\cite{DT074,DT077}. Using $+3e$ as the net
charge of the enclosed La and $3.5$~{\AA} as the radius of the
C$_{60}$ shell, we estimated in this way an energy gain of
$1.7$~eV, the same order of magnitude as the value obtained in the
DFT calculation.

% Hydrogenation
If the reason for altering the energetic preference from an even
to an odd number of attached CF$_3$ radicals is leaving or not
leaving a C radical behind, then a single chemisorbed H atom
chemisorbed on the same double-bond as a lone CF$_3$ radical may
reverse this preference. To investigate this possibility, we
studied the structural and energetic changes associated with the
chemisorption of a hydrogen atom on bare and CF$_3$-functionalized
fullerenes and metallofullerenes and present our results in
Figure~\ref{fig3}.

%===========< FIGURE 4 >=========================================
\begin{figure*}[t]
\centering
\includegraphics[width=2.0\columnwidth]{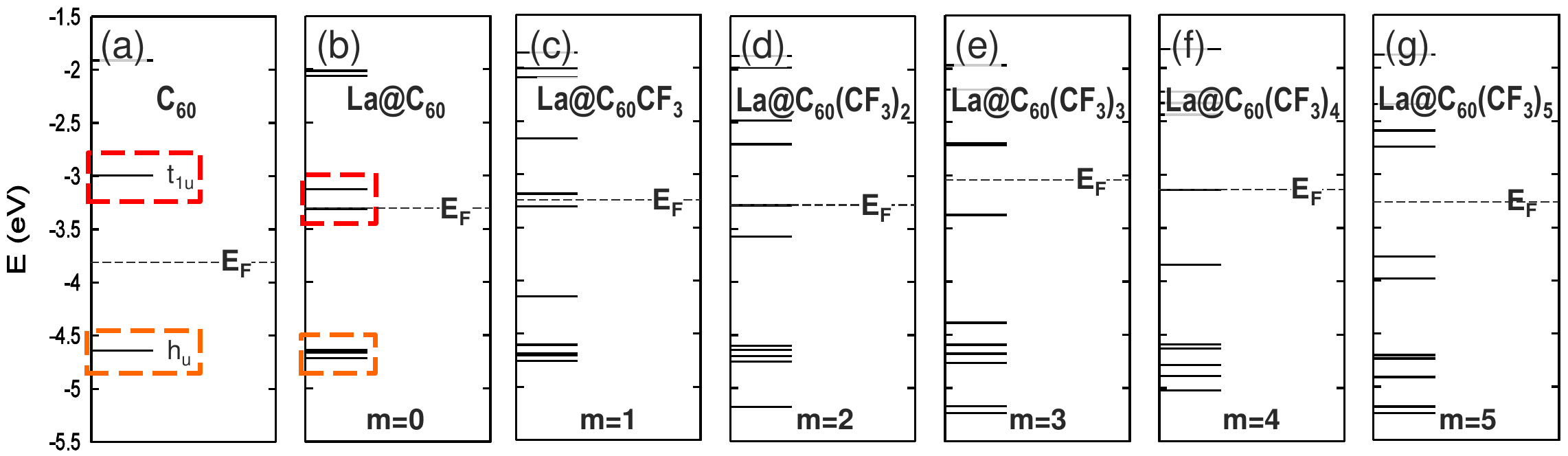}
\caption{(Color online) DFT-PBE molecular orbital energies of (a)
the bare C$_{60}$ molecule, (b) the bare La@C$_{60}$
metallofullerene, and La@C$_{60}$(CF$_3$)$_m$ functionalized with
(c) $m=1$, (d) $m=2$, (e) $m=3$, (f) $m=4$, (g) $m=5$ CF$_3$
radicals. The corresponding (degenerate) energies of the highest
occupied and lowest unoccupied states of C$_{60}$ and La@C$_{60}$
are highlighted in (a) and (b). %
\label{fig4} }
\end{figure*}
%===========< FIGURE 4 >=========================================

We defined the energy change ${\Delta}E^H$ associated with
attaching a single H atom to a bare of CF$_3$-functionalized
fullerene C$_{60}$ or metallofullerene La@C$_{60}$, denoted by R,
by
\begin{equation}
{\Delta}E^H = %
E_{tot}{\rm (H/R)} %
- E_{tot}{\rm (R)} %
- \frac{1}{2}E_{tot}{\rm (H}_2{\rm )} \,. %
\label{eq2} %
\end{equation}
A negative value of ${\Delta}E^H$ indicates that the adsorption of
an H atom, which had initially been formed by dissociating an
H$_2$ molecule, is exothermic.

The top row in Figure~\ref{fig3} describes different ways of
chemically functionalizing a C$_{60}$ cage. The positive value of
${\Delta}E^H$ in the top panel of Figure~\ref{fig3}(a) indicates
that hydrogen prefers not to adsorb on the bare C$_{60}$ cage, as
discussed previously for carbon nanotubes and
fullerenes~\cite{DT193,DT200}. This behavior changes in presence
of an adsorbed CF$_3$ radical, which modifies the electronic
structure of the cage. Noting that adsorption of an isolated H
atom or an isolated CF$_3$ radical on the cage are both
endothermic processes, it is remarkable that the co-adsorption of
CF$_3$ and H turns strongly exothermic, as seen in the top panels
of Figs.~\ref{fig3}(b) and \ref{fig3}(c). As anticipated above and
seen in the top panel of Figure~\ref{fig3}(c), the largest energy
gain occurs when the adsorbates attach to the same double-bond on
the cage, in ortho arrangement on the same hexagonal ring,
changing the configuration of the adjacent C atoms from $sp^2$ to
$sp^3$ and leaving no C radical behind. Somewhat less favorable is
co-adsorption in a para arrangement on the same hexagon, shown in
the top panel of Figure~\ref{fig3}(b). The adsorption sites and
the differences in local bonding mentioned above are best
discussed using the Schlegel diagrams in the bottom panels of
Figure~\ref{fig3}.

The energetics of chemical functionalization changes completely
when a La atom becomes enclosed in the C$_{60}$ cage. Since the
adsorption of an isolated H atom or an isolated CF$_3$ radical
become exothermic processes on La@C$_{60}$, it is not surprising
that all values of ${\Delta}E^H$ are negative, as seen in the
second row of Figure~\ref{fig3}. This means, in other words, that
hydrogenation of La@C$_{60}$(CF$_3$)$_m$ is always exothermic,
irrespective of the presence of CF$_3$ radicals on the surface.
Unlike in the top row, the three values of ${\Delta}E^H$ in the
second row are much more similar: adsorbing an extra CF$_3$
radical next to a pre-adsorbed H atom lowers the energy by
${\approx}0.1$~eV in the para configuration and raises it by
${\approx}0.2$~eV in the ortho configuration. We note the small
energetic preference for the para configuration in this case,
which is reminiscent of the most stable regioisomer of
La@C$_{60}$(CF$_3$)$_2$ shown in Figure~\ref{fig1}(c).

The change in behavior can be attributed to the 3 electrons
donated to the C$_{60}$ cage from the encapsulated La atom. Each
of these extra electrons can be partly transferred to the
electronegative CF$_3$ radical and participate in bonding, leaving
no C radical behind. This is the reason behind our finding in
Figure~\ref{fig2} that La@C$_{60}$(CF$_3$)$_m$ is most stable for
$m=3$.

%Electronic structure

As suggested by the above results, electronic structure plays the
key role in determining both the stability and reactivity of
fullerenes and (functionalized) metallofullerenes. The electronic
eigenstates of the optimized bare C$_{60}$ molecule and of the
La@C$_{60}$(CF$_3$)$_m$ molecules with $0{\leq}m{\leq}5$, depicted
in Figure~\ref{fig1}, are presented in Figure~\ref{fig4}. The
energy eigenvalues are the Kohn-Sham energies of our DFT-PBE
calculations, which are known to underestimate the HOMO-LUMO gap.

As seen in Figure~\ref{fig4}(a), the calculated gap between the
fivefold degenerate $h_u$ HOMO and the threefold degenerate
$t_{1u}$ LUMO of the highly symmetric C$_{60}$ molecule is
${\approx}1.6$~eV wide, slightly lower than the observed
value~\cite{C60gap} of $1.9$~eV. The calculated electronic
spectrum of La@C$_{60}$ is shown in Figure~\ref{fig4}(b). As
expected, the off-center displacement of the La atom reduces the
symmetry and the degeneracy of electronic states. For the sake of
simple comparison, we highlighted the occupied $h_u$ state of
C$_{60}$ and the corresponding occupied eigenstates in La@C$_{60}$
by the orange-bounded rectangle. As seen in Figure~\ref{fig4}(b),
the symmetry lowering due to the off-center displacement of La
splits the $h_u$ state of C$_{60}$ into three closely-spaced
states, two of which are doubly degenerate. More interesting is
the behavior of the initially unfilled $t_{1u}$ state in the
C$_{60}$ molecule, which splits into two states, as highlighted by
the red-bounded rectangle in Figs.~\ref{fig4}(a) and
\ref{fig4}(b). The lower of these states in La@C$_{60}$ is doubly
degenerate and acquires 3 extra electrons from the enclosed La
atom. Consequently, this partly occupied state defines the Fermi
level and provides a metallic character for the La@C$_{60}$
molecule. The significant fundamental band gap is responsible for
the high stability and low reactivity of the C$_{60}$ molecule,
whereas a vanishing band gap
renders La@C$_{60}$ highly reactive. %
To better understand the effect of the encapsuled La atom on the
electronic structure of the La@C$_{60}$ molecule, we discuss the
frontier states near the Fermi level of C$_{60}$ and La@C$_{60}$
molecules in the Supporting Information. %

For La@C$_{60}$(CF$_3$)$_m$ metallofullerenes functionalized with
$m{\geq}1$ CF$_3$ radicals, the eigenstates undergo a complex
change and can no longer be simply compared to those of the bare
C$_{60}$ molecule. Comparing our results in Figs.~\ref{fig4}(b-g),
we find zero fundamental band gap in molecules with an even number
$m$ of CF$_3$ radicals and a nonzero band gap if $m$ is odd. This
finding suggests that La@C$_{60}$ molecules with an odd number of
CF$_3$ radicals should be more stable, in accord with our previous
findings based on total energy calculations. Among
La@C$_{60}$(CF$_3$)$_m$ molecules, we found the largest HOMO-LUMO
gap $E_g{\approx}1.0$~eV for $m=5$, followed by
$E_g{\approx}0.7$~eV for $m=3$ and a very small value
$E_g{\approx}0.1$~eV for $m=1$. Consequently, we expect
metallofullerenes with $m=3$ and $m=5$ CF$_3$ radicals to be most
abundant and easiest to separate from the primary soot, as also
suggested by preliminary experimental
data~\cite{Shinohara-private}. Metallofullerenes with an even
number of CF$_3$ radicals have unpaired electrons at the Fermi
level, which makes these molecules more reactive and less stable.

All molecules discussed above will be further stabilized when
crystallizing in a lattice. We performed corresponding
calculations and found that, same as C$_{60}$ molecules, the
metallofullerenes prefer the close-packed fcc lattice, and also
found no indication of dimerization in the lattice. One issue that
is typically neglected when judging the stability of crystalline
lattices using single-molecule description is the effect of the
long-range dipole-dipole interaction. To judge the importance of
this effect, we performed DFT-PBE calculations for selected
fullerene and metallofullerene lattices. Even though the PBE
exchange-correlation functional does not describe the van der
Waals interaction accurately, such calculations still provide an
adequate description of the charge redistribution and
dipole-dipole interaction. C$_{60}$ molecules, carrying no dipole
moment, crystallize preferably in an fcc lattice with a
nearest-neighbor distance of 10.83~{\AA}, gaining energetically
${\approx}0.06$~eV per molecule.
% Delta E = 0.064~eV/molecule
According to our Bader charge analysis, La@C$_{60}$
metallofullerenes carry a dipole moment $p=3.3$~D
% La@C$_{60}$: p = 3.34 D
and crystallize in an fcc lattice with a slightly reduced
nearest-neighbor distance of 10.72~{\AA}, gaining
${\approx}0.10$~eV per molecule.
% Delta E = 0.096~eV/molecule
The stabilization of the La@C$_{60}$ lattice with respect to
crystalline C$_{60}$ lattice, however small, benefits to a large
degree from the long-range dipole-dipole interaction, since an fcc
lattice of $p=3.3$~D point dipoles should be stabilized by a
comparable energy of ${\approx}0.06$~eV/molecule.
% 0.059~eV/molecule
We have furthermore found that functionalization of the
La@C$_{60}$ metallofullerene by CF$_3$ radicals increases the
molecular dipole moment, but also increases the nearest-neighbor
distance.
% La@C$_{60}$(CF$_3$):     p = 12.37 D
% La@C$_{60}$(CF$_3$)$_2$: p = 18.92 D most stable isomer 2(2)
In the fcc lattice of La@C$_{60}$CF$_3$ molecules with a dipole
moment $p{\approx}12.4$~D, the nearest-neighbor distance increases
to $11.72$~{\AA} and the stabilization in the lattice changes to
${\approx}0.09$~eV/molecule.
% 0.093 eV/molecule
Similarly, in the fcc lattice of the most stable
La@C$_{60}$(CF$_3$)$_2$ molecules with a dipole moment
$p{\approx}18.9$~D, the nearest-neighbor distance changes to
$11.68$~{\AA} and the stabilization in the lattice changes to
${\approx}0.10$~eV/molecule.
% 0.099 eV/molecule
The above DFT-PBE values of the lattice stabilization energy in
CF$_3$-functionalized metallofullerenes are one order of magnitude
smaller than estimates based on the interaction of point dipoles.
Our overall finding is that long-range dipole-dipole interaction
is of secondary importance for these lattices.
% La@C$_{60}$CF$_3$ fcc: -0.663 eV/molecule when using point dipoles
% La@C$_{60}$(CF$_3$)$_2$ fcc: -1.576 eV/molecule with point dipoles

\section{Conclusions}

In summary, exploring the possibility to obtain superconducting
behavior in crystalline La@C$_{60}$ even under harsh ambient
conditions, we provided theoretical support for the experimental
effort to solubilize the La@C$_{60}$ endohedral metallofullerene
from the primary soot. We used {\em ab initio} density functional
theory calculations to compare the stability and electronic
structure of the bare C$_{60}$ fullerene and La@C$_{60}$ to their
counterparts functionalized by $m$ CF$_3$ radicals. We found that
several CF$_3$ radicals can form stable bonds to C$_{60}$ and
La@C$_{60}$, and have identified the most stable structural
isomers for $m{\leq}5$. Generally, structures with an even $m$ are
energetically preferred for C$_{60}$ and structures with an odd
$m$ for La@C$_{60}$ due to the extra 3 electrons donated by the
encapsulated La. This is consistent with our finding that a wide
HOMO-LUMO gap opens in La@C$_{60}$(CF$_3$)$_m$ molecules with an
odd $m$ value, causing extra stabilization in the closed-shell
electronic configuration. %
We also addressed the possibility of a single hydrogen atom
adsorbing at a C radical site in La@C$_{60}$(CF$_3$)$_m$, which
would further stabilize the open-shell configuration for odd $m$
values and reduce the even/odd alternation in stability. %
We found that the two-fold function of CF$_3$ radicals as
stabilizing agents and molecular separators may help in isolating
specific molecules from the primary soot. If
La@C$_{60}$(CF$_3$)$_m$ could be solubilized and recrystallized,
endohedral instead of exohedral doping by three electrons per
C$_{60}$ and use of stabilizing CF$_3$ radicals as spacers may
lead to superconductivity with a relatively high $T_c$ achieved by
optimizing the lattice constant of the molecular crystal.
Stability even under harsh ambient conditions should then be
expected due to the absence of alkali atoms filling the
interstitial sites in known M$_3$C$_{60}$ superconductors.

\section{Computational Techniques}

We utilized {\em ab initio} density functional theory (DFT) as
implemented in the \textsc{VASP} code~\cite{VASP,VASP2,VASP3} to
optimize the structure of C$_{60}$ and La@C$_{60}$(CF$_3$)$_m$
functionalized with $0{\leq}m{\leq}5$ CF$_3$ radicals and obtained
the total energy as well as the electronic structure for these
systems. We used projector-augmented-wave (PAW)
pseudopotentials~\cite{PAWPseudo} and the Perdew-Burke-Ernzerhof
(PBE)~\cite{PBE} exchange-correlation functional. All isolated
structures have been represented using periodic boundary
conditions and separated by a $12$~{\AA} thick vacuum region, so
that no band dispersion could be detected. We used $500$~eV as the
electronic kinetic energy cutoff for the plane-wave basis and a
total energy difference between subsequent iterations below
$10^{-5}$~eV as the criterion for reaching self-consistency. All
geometries have been optimized using the conjugate-gradient
method~\cite{CGmethod} until none of the residual Hellmann-Feynman
forces exceeded $10^{-2}$~eV/{\AA}.

\begin{suppinfo}
Additional information is provided regarding the origin of
superconductivity in alkali-based M$_3$C$_{60}$ solids, charge
distribution in the frontier states of pristine C$_{60}$ and
La@C$_{60}$ molecules, and the equilibrium structure and
stability of different La@C$_{60}$(CF$_3$)$_m$ regioisomers.\\
\end{suppinfo}
\quad\par
%%%%%%%%%%%%%%%%%%%%%%%%%%%%%%%%%%%%%%%%%%%%%%%%%%%%%%%%%%%%%%%%%%%%%
% Author information
%%%%%%%%%%%%%%%%%%%%%%%%%%%%%%%%%%%%%%%%%%%%%%%%%%%%%%%%%%%%%%%%%%%%%

{\noindent\bf Author Information}\\

{\noindent\bf Corresponding Author}\\
$^*$E-mail: {\tt tomanek@pa.msu.edu} \\

{\noindent\bf Notes}\\
The authors declare no competing financial interest.

%%%%%%%%%%%%%%%%%%%%%%%%%%%%%%%%%%%%%%%%%%%%%%%%%%%%%%%%%%%%%%%%%%%%%
%% Acknowledgements should use the acknowledgement environment.
%%%%%%%%%%%%%%%%%%%%%%%%%%%%%%%%%%%%%%%%%%%%%%%%%%%%%%%%%%%%%%%%%%%%%
\begin{acknowledgement}
We would like to acknowledge useful discussions with Hisanori
Shinohara, who inspired this study and advised us of experimental
progress in his group, and with Zhongqi Jin. This study was
supported by the NSF/AFOSR EFRI 2-DARE grant number
\#EFMA-1433459. Computational resources have been provided by the
Michigan State University High Performance Computing Center.
\end{acknowledgement}

%\bibliography{MF17}

\begin{mcitethebibliography}{49}
\providecommand*\natexlab[1]{#1}
\providecommand*\mciteSetBstSublistMode[1]{}
\providecommand*\mciteSetBstMaxWidthForm[2]{}
\providecommand*\mciteBstWouldAddEndPuncttrue
  {\def\EndOfBibitem{\unskip.}}
\providecommand*\mciteBstWouldAddEndPunctfalse
  {\let\EndOfBibitem\relax}
\providecommand*\mciteSetBstMidEndSepPunct[3]{}
\providecommand*\mciteSetBstSublistLabelBeginEnd[3]{}
\providecommand*\EndOfBibitem{} \mciteSetBstSublistMode{f}
\mciteSetBstMaxWidthForm{subitem}{(\alph{mcitesubitemcount})}
\mciteSetBstSublistLabelBeginEnd
  {\mcitemaxwidthsubitemform\space}
  {\relax}
  {\relax}

\bibitem[Kroto \latin{et~al.}(1985)Kroto, Heath, O'Brien, Curl, and
  Smalley]{C60Kroto}
Kroto,~H.~W.; Heath,~J.~R.; O'Brien,~S.~C.; Curl,~R.~F.;
Smalley,~R.~E.
  \emph{Nature} \textbf{1985}, \emph{318}, 162--163\relax
\mciteBstWouldAddEndPuncttrue
\mciteSetBstMidEndSepPunct{\mcitedefaultmidpunct}
{\mcitedefaultendpunct}{\mcitedefaultseppunct}\relax \EndOfBibitem
\bibitem[Heath \latin{et~al.}(1985)Heath, O'Brien, Zhang, Liu, Curl, Tittel,
  and Smalley]{Heath1985}
Heath,~J.~R.; O'Brien,~S.~C.; Zhang,~Q.; Liu,~Y.; Curl,~R.~F.;
Tittel,~F.~K.;
  Smalley,~R.~E. \emph{J. Am. Chem. Soc.} \textbf{1985}, \emph{107},
  7779--7780\relax
\mciteBstWouldAddEndPuncttrue
\mciteSetBstMidEndSepPunct{\mcitedefaultmidpunct}
{\mcitedefaultendpunct}{\mcitedefaultseppunct}\relax \EndOfBibitem
\bibitem[Chai \latin{et~al.}(1991)Chai, Guo, Jin, Haufler, Chibante, Fure,
  Wang, Alford, and Smalley]{Chai1991}
Chai,~Y.; Guo,~T.; Jin,~C.; Haufler,~R.~E.; Chibante,~L. P.~F.;
Fure,~J.;
  Wang,~L.; Alford,~J.~M.; Smalley,~R.~E. \emph{J. Phys. Chem.} \textbf{1991},
  \emph{95}, 7564--7568\relax
\mciteBstWouldAddEndPuncttrue
\mciteSetBstMidEndSepPunct{\mcitedefaultmidpunct}
{\mcitedefaultendpunct}{\mcitedefaultseppunct}\relax \EndOfBibitem
\bibitem[Haufler \latin{et~al.}(1990)Haufler, Conceicao, Chibante, Chai, Byrne,
  Flanagan, Haley, O'Brien, Pan, Xiao, Billups, Ciufolini, Hauge, Margrave,
  Wilson, Curl, and Smalley]{Haufler90}
Haufler,~R.~E. \latin{et~al.}  \emph{J. Phys. Chem.}
\textbf{1990}, \emph{94},
  8634--8636\relax
\mciteBstWouldAddEndPuncttrue
\mciteSetBstMidEndSepPunct{\mcitedefaultmidpunct}
{\mcitedefaultendpunct}{\mcitedefaultseppunct}\relax \EndOfBibitem
\bibitem[Shinohara \latin{et~al.}(1993)Shinohara, Yamaguchi, Hayashi, Sato,
  Ohkohchi, Ando, and Saito]{Shinohara1993jpc}
Shinohara,~H.; Yamaguchi,~H.; Hayashi,~N.; Sato,~H.; Ohkohchi,~M.;
Ando,~Y.;
  Saito,~Y. \emph{J. Phys. Chem.} \textbf{1993}, \emph{97}, 4259--4261\relax
\mciteBstWouldAddEndPuncttrue
\mciteSetBstMidEndSepPunct{\mcitedefaultmidpunct}
{\mcitedefaultendpunct}{\mcitedefaultseppunct}\relax \EndOfBibitem
\bibitem[Tellgmann \latin{et~al.}(1996)Tellgmann, Krawez, Lin, Hertel, and
  Campbell]{Tellgmann96}
Tellgmann,~R.; Krawez,~N.; Lin,~S.-H.; Hertel,~I.~V.; Campbell,~E.
E.~B.
  \emph{Nature} \textbf{1996}, \emph{382}, 407--408\relax
\mciteBstWouldAddEndPuncttrue
\mciteSetBstMidEndSepPunct{\mcitedefaultmidpunct}
{\mcitedefaultendpunct}{\mcitedefaultseppunct}\relax \EndOfBibitem
\bibitem[Campbell \latin{et~al.}(1997)Campbell, Tellgmann, Krawez, and
  Hertel]{CAMPBELL19971763}
Campbell,~E.; Tellgmann,~R.; Krawez,~N.; Hertel,~I. \emph{J. Phys.
Chem.
  Solids} \textbf{1997}, \emph{58}, 1763--1769\relax
\mciteBstWouldAddEndPuncttrue
\mciteSetBstMidEndSepPunct{\mcitedefaultmidpunct}
{\mcitedefaultendpunct}{\mcitedefaultseppunct}\relax \EndOfBibitem
\bibitem[Almeida~Murphy \latin{et~al.}(1996)Almeida~Murphy, Pawlik, Weidinger,
  H\"ohne, Alcala, and Spaeth]{Almeida96}
Almeida~Murphy,~T.; Pawlik,~T.; Weidinger,~A.; H\"ohne,~M.;
Alcala,~R.;
  Spaeth,~J.-M. \emph{Phys. Rev. Lett.} \textbf{1996}, \emph{77},
  1075--1078\relax
\mciteBstWouldAddEndPuncttrue
\mciteSetBstMidEndSepPunct{\mcitedefaultmidpunct}
{\mcitedefaultendpunct}{\mcitedefaultseppunct}\relax \EndOfBibitem
\bibitem[Knapp \latin{et~al.}(1997)Knapp, Dinse, Pietzak, Waiblinger, and
  Weidinger]{KNAPP1997433}
Knapp,~C.; Dinse,~K.-P.; Pietzak,~B.; Waiblinger,~M.;
Weidinger,~A. \emph{Chem.
  Phys. Lett.} \textbf{1997}, \emph{272}, 433--437\relax
\mciteBstWouldAddEndPuncttrue
\mciteSetBstMidEndSepPunct{\mcitedefaultmidpunct}
{\mcitedefaultendpunct}{\mcitedefaultseppunct}\relax \EndOfBibitem
\bibitem[Mauser \latin{et~al.}(1997)Mauser, Hirsch, van Eikema~Hommes, Clark,
  Pietzak, Weidinger, and Dunsch]{Mauser97}
Mauser,~H.; Hirsch,~A.; van Eikema~Hommes,~N. J.~R.; Clark,~T.;
Pietzak,~B.;
  Weidinger,~A.; Dunsch,~L. \emph{Angew. Chem. Int. Ed.} \textbf{1997},
  \emph{36}, 2835--2838\relax
\mciteBstWouldAddEndPuncttrue
\mciteSetBstMidEndSepPunct{\mcitedefaultmidpunct}
{\mcitedefaultendpunct}{\mcitedefaultseppunct}\relax \EndOfBibitem
\bibitem[Hirata \latin{et~al.}(1996)Hirata, Hatakeyama, Mieno, and
  Sato]{Hirata96}
Hirata,~T.; Hatakeyama,~R.; Mieno,~T.; Sato,~N. \emph{J. Vac. Sci.
Technol. A}
  \textbf{1996}, \emph{14}, 615--618\relax
\mciteBstWouldAddEndPuncttrue
\mciteSetBstMidEndSepPunct{\mcitedefaultmidpunct}
{\mcitedefaultendpunct}{\mcitedefaultseppunct}\relax \EndOfBibitem
\bibitem[Weck \latin{et~al.}(2010)Weck, Kim, Czerwinski, and
  Tom\'anek]{Weck-TcC60}
Weck,~P.~F.; Kim,~E.; Czerwinski,~K.~R.; Tom\'anek,~D. \emph{Phys.
Rev. B}
  \textbf{2010}, \emph{81}, 125448\relax
\mciteBstWouldAddEndPuncttrue
\mciteSetBstMidEndSepPunct{\mcitedefaultmidpunct}
{\mcitedefaultendpunct}{\mcitedefaultseppunct}\relax \EndOfBibitem
\bibitem[Shinohara(2000)]{Shinohara2000}
Shinohara,~H. \emph{Rep. Prog. Phys.} \textbf{2000}, \emph{63},
843\relax \mciteBstWouldAddEndPuncttrue
\mciteSetBstMidEndSepPunct{\mcitedefaultmidpunct}
{\mcitedefaultendpunct}{\mcitedefaultseppunct}\relax \EndOfBibitem
\bibitem[Shi()]{Shinohara-private}
H. Shinohara (private communication).\relax
\mciteBstWouldAddEndPunctfalse
\mciteSetBstMidEndSepPunct{\mcitedefaultmidpunct}
{}{\mcitedefaultseppunct}\relax \EndOfBibitem
\bibitem[Schluter \latin{et~al.}(1992)Schluter, Lannoo, Needels, Baraff, and
  Tomanek]{DT053}
Schluter,~M.; Lannoo,~M.; Needels,~M.; Baraff,~G.~A.; Tomanek,~D.
\emph{Phys.
  Rev. Lett.} \textbf{1992}, \emph{68}, 526--529\relax
\mciteBstWouldAddEndPuncttrue
\mciteSetBstMidEndSepPunct{\mcitedefaultmidpunct}
{\mcitedefaultendpunct}{\mcitedefaultseppunct}\relax \EndOfBibitem
\bibitem[Schluter \latin{et~al.}(1992)Schluter, Lannoo, Needels, Baraff, and
  Tomanek]{DT057}
Schluter,~M.; Lannoo,~M.; Needels,~M.; Baraff,~G.~A.; Tomanek,~D.
\emph{J.
  Phys. Chem. Solids} \textbf{1992}, \emph{53}, 1473--1485\relax
\mciteBstWouldAddEndPuncttrue
\mciteSetBstMidEndSepPunct{\mcitedefaultmidpunct}
{\mcitedefaultendpunct}{\mcitedefaultseppunct}\relax \EndOfBibitem
\bibitem[Yamamoto \latin{et~al.}(1994)Yamamoto, Funasaka, Takahashi, and
  Akasaka]{Yamamoto1994}
Yamamoto,~K.; Funasaka,~H.; Takahashi,~T.; Akasaka,~T. \emph{J.
Phys. Chem.}
  \textbf{1994}, \emph{98}, 2008--2011\relax
\mciteBstWouldAddEndPuncttrue
\mciteSetBstMidEndSepPunct{\mcitedefaultmidpunct}
{\mcitedefaultendpunct}{\mcitedefaultseppunct}\relax \EndOfBibitem
\bibitem[Kikuchi \latin{et~al.}(1993)Kikuchi, Suzuki, Nakao, Nakahara,
  Wakabayashi, Shiromaru, Saito, Ikemoto, and Achiba]{KIKUCHI199367}
Kikuchi,~K.; Suzuki,~S.; Nakao,~Y.; Nakahara,~N.; Wakabayashi,~T.;
  Shiromaru,~H.; Saito,~K.; Ikemoto,~I.; Achiba,~Y. \emph{Chem. Phys. Lett.}
  \textbf{1993}, \emph{216}, 67--71\relax
\mciteBstWouldAddEndPuncttrue
\mciteSetBstMidEndSepPunct{\mcitedefaultmidpunct}
{\mcitedefaultendpunct}{\mcitedefaultseppunct}\relax \EndOfBibitem
\bibitem[Wang \latin{et~al.}(2013)Wang, Nakanishi, Noda, Niwa, Zhang, Kitaura,
  and Shinohara]{wang2013}
Wang,~Z.; Nakanishi,~Y.; Noda,~S.; Niwa,~H.; Zhang,~J.;
Kitaura,~R.;
  Shinohara,~H. \emph{Angew. Chem. Int. Ed.} \textbf{2013}, \emph{52},
  11770--11774\relax
\mciteBstWouldAddEndPuncttrue
\mciteSetBstMidEndSepPunct{\mcitedefaultmidpunct}
{\mcitedefaultendpunct}{\mcitedefaultseppunct}\relax \EndOfBibitem
\bibitem[Ogawa \latin{et~al.}(2000)Ogawa, Sugai, , and Shinohara]{Ogawa2000}
Ogawa,~T.; Sugai,~T.; ; Shinohara,~H. \emph{J. Am. Chem. Soc.}
\textbf{2000},
  \emph{122}, 3538--3539\relax
\mciteBstWouldAddEndPuncttrue
\mciteSetBstMidEndSepPunct{\mcitedefaultmidpunct}
{\mcitedefaultendpunct}{\mcitedefaultseppunct}\relax \EndOfBibitem
\bibitem[Kubozono \latin{et~al.}(1996)Kubozono, Noto, Ohta, Maeda, Kashino,
  Emura, Ukita, and Sogabe]{Kubozono96}
Kubozono,~Y.; Noto,~T.; Ohta,~T.; Maeda,~H.; Kashino,~S.;
Emura,~S.; Ukita,~S.;
  Sogabe,~T. \emph{Chem. Lett.} \textbf{1996}, \emph{25}, 453--454\relax
\mciteBstWouldAddEndPuncttrue
\mciteSetBstMidEndSepPunct{\mcitedefaultmidpunct}
{\mcitedefaultendpunct}{\mcitedefaultseppunct}\relax \EndOfBibitem
\bibitem[Wang \latin{et~al.}(1993)Wang, Alford, Chai, Diener, Zhang, McClure,
  Guo, Scuseria, and Smalley]{WANG1993354}
Wang,~L.; Alford,~J.; Chai,~Y.; Diener,~M.; Zhang,~J.;
McClure,~S.; Guo,~T.;
  Scuseria,~G.; Smalley,~R. \emph{Chem. Phys. Lett.} \textbf{1993}, \emph{207},
  354--359\relax
\mciteBstWouldAddEndPuncttrue
\mciteSetBstMidEndSepPunct{\mcitedefaultmidpunct}
{\mcitedefaultendpunct}{\mcitedefaultseppunct}\relax \EndOfBibitem
\bibitem[Wang \latin{et~al.}(1993)Wang, Alford, Chai, Diener, and
  Smalley]{Wang1993}
Wang,~L.-S.; Alford,~J.~M.; Chai,~Y.; Diener,~M.; Smalley,~R.~E.
\emph{Z. Phys.
  D.} \textbf{1993}, \emph{26}, 297--299\relax
\mciteBstWouldAddEndPuncttrue
\mciteSetBstMidEndSepPunct{\mcitedefaultmidpunct}
{\mcitedefaultendpunct}{\mcitedefaultseppunct}\relax \EndOfBibitem
\bibitem[Kubozono \latin{et~al.}(1995)Kubozono, Ohta, Hayashibara, Maeda,
  Ishida, Kashino, Oshima, Yamazaki, Ukita, and Sogabe]{Kubozono1995}
Kubozono,~Y.; Ohta,~T.; Hayashibara,~T.; Maeda,~H.; Ishida,~H.;
Kashino,~S.;
  Oshima,~K.; Yamazaki,~H.; Ukita,~S.; Sogabe,~T. \emph{Chem. Lett.}
  \textbf{1995}, \emph{24}, 457--458\relax
\mciteBstWouldAddEndPuncttrue
\mciteSetBstMidEndSepPunct{\mcitedefaultmidpunct}
{\mcitedefaultendpunct}{\mcitedefaultseppunct}\relax \EndOfBibitem
\bibitem[Wang \latin{et~al.}(2016)Wang, Aoyagi, Omachi, Kitaura, and
  Shinohara]{wang2016}
Wang,~Z.; Aoyagi,~S.; Omachi,~H.; Kitaura,~R.; Shinohara,~H.
\emph{Angew. Chem.
  Int. Ed.} \textbf{2016}, \emph{55}, 199--202\relax
\mciteBstWouldAddEndPuncttrue
\mciteSetBstMidEndSepPunct{\mcitedefaultmidpunct}
{\mcitedefaultendpunct}{\mcitedefaultseppunct}\relax \EndOfBibitem
\bibitem[Diener and Alford(1998)Diener, and Alford]{Diener1998}
Diener,~M.~D.; Alford,~J.~M. \emph{Nature} \textbf{1998},
\emph{393},
  668--671\relax
\mciteBstWouldAddEndPuncttrue
\mciteSetBstMidEndSepPunct{\mcitedefaultmidpunct}
{\mcitedefaultendpunct}{\mcitedefaultseppunct}\relax \EndOfBibitem
\bibitem[Ganin \latin{et~al.}(2010)Ganin, Takabayashi, Jegli\v{c}, Ar\v{c}on,
  Poto\v{c}nik, Baker, Ohishi, McDonald, Tzirakis, McLennan, Darling, Takata,
  Rosseinsky, and Prassides]{Ganin-10}
  Ganin,~A.~Y.; Takabayashi,~Y.; Jegli\v{c},~P.; Ar\v{c}on,~D.; Poto\v{c}nik,~A.;
  Baker,~P.~J.; Ohishi,~Y.; McDonald,~M.~T.; Tzirakis,~M.~D.;
  McLennan,~A.; Darling, G. R.; Takata, M.; Rosseinsky, M. J.; Prassides, K.
  Polymorphism control of superconductivity and magnetism in
  Cs$_3$C$_{60}$ close to the Mott transition. \emph{Nature} \textbf{2010},
  \emph{466}, 221--225\relax
\mciteBstWouldAddEndPuncttrue
\mciteSetBstMidEndSepPunct{\mcitedefaultmidpunct}
{\mcitedefaultendpunct}{\mcitedefaultseppunct}\relax \EndOfBibitem
\bibitem[Nomura \latin{et~al.}(2015)Nomura, Sakai, Capone, and
  Arita]{Nomura-eC60-15}
Nomura,~Y.; Sakai,~S.; Capone,~M.; Arita,~R. \emph{Science Adv.}
\textbf{2015},
  \emph{1}, e1500568\relax
\mciteBstWouldAddEndPuncttrue
\mciteSetBstMidEndSepPunct{\mcitedefaultmidpunct}
{\mcitedefaultendpunct}{\mcitedefaultseppunct}\relax \EndOfBibitem
\bibitem[Takeda \latin{et~al.}(2006)Takeda, Yokoyama, Ito, Miyazaki, Shimotani,
  Yakigaya, Kakiuchi, Sawa, Takagi, Kitazawa, and Dragoe]{Takeda-ArC60}
Takeda,~A.; Yokoyama,~Y.; Ito,~S.; Miyazaki,~T.; Shimotani,~H.;
Yakigaya,~K.;
  Kakiuchi,~T.; Sawa,~H.; Takagi,~H.; Kitazawa,~K.; Dragoe,~N. \emph{Chem.
  Commun.} \textbf{2006}, 912--914\relax
\mciteBstWouldAddEndPuncttrue
\mciteSetBstMidEndSepPunct{\mcitedefaultmidpunct}
{\mcitedefaultendpunct}{\mcitedefaultseppunct}\relax \EndOfBibitem
\bibitem[Perdew \latin{et~al.}(1996)Perdew, Burke, and Ernzerhof]{PBE}
Perdew,~J.~P.; Burke,~K.; Ernzerhof,~M. \emph{Phys. Rev. Lett.}
\textbf{1996},
  \emph{77}, 3865--3868\relax
\mciteBstWouldAddEndPuncttrue
\mciteSetBstMidEndSepPunct{\mcitedefaultmidpunct}
{\mcitedefaultendpunct}{\mcitedefaultseppunct}\relax \EndOfBibitem
\bibitem[Darwish \latin{et~al.}()Darwish, Abdul-Sada, Avent, Martsinovich,
  Street, and Taylor]{Darwish2004}
Darwish,~A.~D.; Abdul-Sada,~A.~K.; Avent,~A.~G.; Martsinovich,~N.;
  Street,~J.~M.; Taylor,~R. \relax
\mciteBstWouldAddEndPunctfalse
\mciteSetBstMidEndSepPunct{\mcitedefaultmidpunct}
{}{\mcitedefaultseppunct}\relax \EndOfBibitem
\bibitem[Dorozhkin \latin{et~al.}(2007)Dorozhkin, Goryunkov, Ioffe, Avdoshenko,
  Markov, Tamm, Ignat'eva, Sidorov, and Troyanov]{trifluo2007}
Dorozhkin,~E.~I.; Goryunkov,~A.~A.; Ioffe,~I.~N.;
Avdoshenko,~S.~M.;
  Markov,~V.~Y.; Tamm,~N.~B.; Ignat'eva,~D.~V.; Sidorov,~L.~N.; Troyanov,~S.~I.
  \emph{Eur. J. Org. Chem.} \textbf{2007}, \emph{2007}, 5082--5094\relax
\mciteBstWouldAddEndPuncttrue
\mciteSetBstMidEndSepPunct{\mcitedefaultmidpunct}
{\mcitedefaultendpunct}{\mcitedefaultseppunct}\relax \EndOfBibitem
\bibitem[Henkelman \latin{et~al.}(2006)Henkelman, Arnaldsson, and
  J\'onsson]{Bader06}
Henkelman,~G.; Arnaldsson,~A.; J\'onsson,~H. \emph{Comput. Mater.
Sci.}
  \textbf{2006}, \emph{36}, 354--360\relax
\mciteBstWouldAddEndPuncttrue
\mciteSetBstMidEndSepPunct{\mcitedefaultmidpunct}
{\mcitedefaultendpunct}{\mcitedefaultseppunct}\relax \EndOfBibitem
\bibitem[Sanville \latin{et~al.}(2007)Sanville, Kenny, Smith, and
  Henkelman]{Bader07}
Sanville,~E.; Kenny,~S.~D.; Smith,~R.; Henkelman,~G. \emph{J.
Comp. Chem.}
  \textbf{2007}, \emph{28}, 899--908\relax
\mciteBstWouldAddEndPuncttrue
\mciteSetBstMidEndSepPunct{\mcitedefaultmidpunct}
{\mcitedefaultendpunct}{\mcitedefaultseppunct}\relax \EndOfBibitem
\bibitem[Tang \latin{et~al.}(2009)Tang, Sanville, and Henkelman]{Bader09}
Tang,~W.; Sanville,~E.; Henkelman,~G. \emph{J. Phys. Condens.
Matter}
  \textbf{2009}, \emph{21}, 084204\relax
\mciteBstWouldAddEndPuncttrue
\mciteSetBstMidEndSepPunct{\mcitedefaultmidpunct}
{\mcitedefaultendpunct}{\mcitedefaultseppunct}\relax \EndOfBibitem
\bibitem[Yu and Trinkle(2011)Yu, and Trinkle]{Bader11}
Yu,~M.; Trinkle,~D.~R. \emph{J. Chem. Phys.} \textbf{2011},
\emph{134},
  064111\relax
\mciteBstWouldAddEndPuncttrue
\mciteSetBstMidEndSepPunct{\mcitedefaultmidpunct}
{\mcitedefaultendpunct}{\mcitedefaultseppunct}\relax \EndOfBibitem
\bibitem[Lee \latin{et~al.}(2009)Lee, Kleis, Rossmeisl, and Morgan]{Lee2009prb}
Lee,~Y.-L.; Kleis,~J.; Rossmeisl,~J.; Morgan,~D. \emph{Phys. Rev.
B}
  \textbf{2009}, \emph{80}, 224101\relax
\mciteBstWouldAddEndPuncttrue
\mciteSetBstMidEndSepPunct{\mcitedefaultmidpunct}
{\mcitedefaultendpunct}{\mcitedefaultseppunct}\relax \EndOfBibitem
\bibitem[Yang \latin{et~al.}(2015)Yang, Dong, Wang, Zhang, Guan, Kuntz, Warren,
  and Tom\'anek]{DT249}
Yang,~T.; Dong,~B.; Wang,~J.; Zhang,~Z.; Guan,~J.; Kuntz,~K.;
Warren,~S.~C.;
  Tom\'anek,~D. \emph{Phys. Rev. B} \textbf{2015}, \emph{92}, 125412\relax
\mciteBstWouldAddEndPuncttrue
\mciteSetBstMidEndSepPunct{\mcitedefaultmidpunct}
{\mcitedefaultendpunct}{\mcitedefaultseppunct}\relax \EndOfBibitem
\bibitem[Wang \latin{et~al.}(1993)Wang, Tomanek, and Ruoff]{DT074}
Wang,~Y.; Tomanek,~D.; Ruoff,~R.~S. \emph{Chem. Phys. Lett.}
\textbf{1993},
  \emph{208}, 79--85\relax
\mciteBstWouldAddEndPuncttrue
\mciteSetBstMidEndSepPunct{\mcitedefaultmidpunct}
{\mcitedefaultendpunct}{\mcitedefaultseppunct}\relax \EndOfBibitem
\bibitem[Li and Tomanek(1994)Li, and Tomanek]{DT077}
Li,~Y.~S.; Tomanek,~D. \emph{Chem. Phys. Lett.} \textbf{1994},
\emph{221},
  453--458\relax
\mciteBstWouldAddEndPuncttrue
\mciteSetBstMidEndSepPunct{\mcitedefaultmidpunct}
{\mcitedefaultendpunct}{\mcitedefaultseppunct}\relax \EndOfBibitem
\bibitem[Miller \latin{et~al.}(2008)Miller, Kintigh, Kim, Weck, Berber, and
  Tomanek]{DT193}
Miller,~G.; Kintigh,~J.; Kim,~E.; Weck,~P.; Berber,~S.;
Tomanek,~D. \emph{J.
  Am. Chem. Soc.} \textbf{2008}, \emph{130}, 2296--2303\relax
\mciteBstWouldAddEndPuncttrue
\mciteSetBstMidEndSepPunct{\mcitedefaultmidpunct}
{\mcitedefaultendpunct}{\mcitedefaultseppunct}\relax \EndOfBibitem
\bibitem[Berber and Tom\'{a}nek(2009)Berber, and Tom\'{a}nek]{DT200}
Berber,~S.; Tom\'{a}nek,~D. \emph{Phys. Rev. B} \textbf{2009},
\emph{80},
  075427\relax
\mciteBstWouldAddEndPuncttrue
\mciteSetBstMidEndSepPunct{\mcitedefaultmidpunct}
{\mcitedefaultendpunct}{\mcitedefaultseppunct}\relax \EndOfBibitem
\bibitem[Weaver \latin{et~al.}(1991)Weaver, Martins, Komeda, Chen, Ohno, Kroll,
  Troullier, Haufler, and Smalley]{C60gap}
Weaver,~J.~H.; Martins,~J.~L.; Komeda,~T.; Chen,~Y.; Ohno,~T.~R.;
Kroll,~G.~H.;
  Troullier,~N.; Haufler,~R.~E.; Smalley,~R.~E. \emph{Phys. Rev. Lett.}
  \textbf{1991}, \emph{66}, 1741--1744\relax
\mciteBstWouldAddEndPuncttrue
\mciteSetBstMidEndSepPunct{\mcitedefaultmidpunct}
{\mcitedefaultendpunct}{\mcitedefaultseppunct}\relax \EndOfBibitem
\bibitem[Kresse and Furthm\"uller(1996)Kresse, and Furthm\"uller]{VASP}
Kresse,~G.; Furthm\"uller,~J. \emph{Phys. Rev. B} \textbf{1996},
\emph{54},
  11169--11186\relax
\mciteBstWouldAddEndPuncttrue
\mciteSetBstMidEndSepPunct{\mcitedefaultmidpunct}
{\mcitedefaultendpunct}{\mcitedefaultseppunct}\relax \EndOfBibitem
\bibitem[Kresse and Hafner(1993)Kresse, and Hafner]{VASP2}
Kresse,~G.; Hafner,~J. \emph{Phys. Rev. B} \textbf{1993},
\emph{47},
  558--561\relax
\mciteBstWouldAddEndPuncttrue
\mciteSetBstMidEndSepPunct{\mcitedefaultmidpunct}
{\mcitedefaultendpunct}{\mcitedefaultseppunct}\relax \EndOfBibitem
\bibitem[Kresse and Hafner(1994)Kresse, and Hafner]{VASP3}
Kresse,~G.; Hafner,~J. \emph{Phys. Rev. B} \textbf{1994},
\emph{49},
  14251--14269\relax
\mciteBstWouldAddEndPuncttrue
\mciteSetBstMidEndSepPunct{\mcitedefaultmidpunct}
{\mcitedefaultendpunct}{\mcitedefaultseppunct}\relax \EndOfBibitem
\bibitem[Kresse and Joubert(1999)Kresse, and Joubert]{PAWPseudo}
Kresse,~G.; Joubert,~D. \emph{Phys. Rev. B} \textbf{1999},
\emph{59},
  1758--1775\relax
\mciteBstWouldAddEndPuncttrue
\mciteSetBstMidEndSepPunct{\mcitedefaultmidpunct}
{\mcitedefaultendpunct}{\mcitedefaultseppunct}\relax \EndOfBibitem
\bibitem[Hestenes and Stiefel(1952)Hestenes, and Stiefel]{CGmethod}
Hestenes,~M.~R.; Stiefel,~E. \emph{J. Res. Natl. Bur. Stand.}
\textbf{1952},
  \emph{49}, 409--436\relax
\mciteBstWouldAddEndPuncttrue
\mciteSetBstMidEndSepPunct{\mcitedefaultmidpunct}
{\mcitedefaultendpunct}{\mcitedefaultseppunct}\relax \EndOfBibitem
\end{mcitethebibliography}
%\end{document}
%%%%%%%%%%%%%%%%%%%%%%%%%%%%%%%%%%%%%%%%%%%%%%%%%%%%%%%%%%%%%%%%%%%%%

\providecommand{\latin}[1]{#1} \makeatletter \providecommand{\doi}
  {\begingroup\let\do\@makeother\dospecials
  \catcode`\{=1 \catcode`\}=2\doi@aux}
\providecommand{\doi@aux}[1]{\endgroup\texttt{#1}} \makeatother
\providecommand*\mcitethebibliography{\thebibliography} \csname
@ifundefined\endcsname{endmcitethebibliography}
  {\let\endmcitethebibliography\endthebibliography}{}

\end{document}

%===========< FIGURE S1 >=============================================
% entire page. There is no need to do explicit centering.
\begin{figure}
\includegraphics[width=1.0\columnwidth]{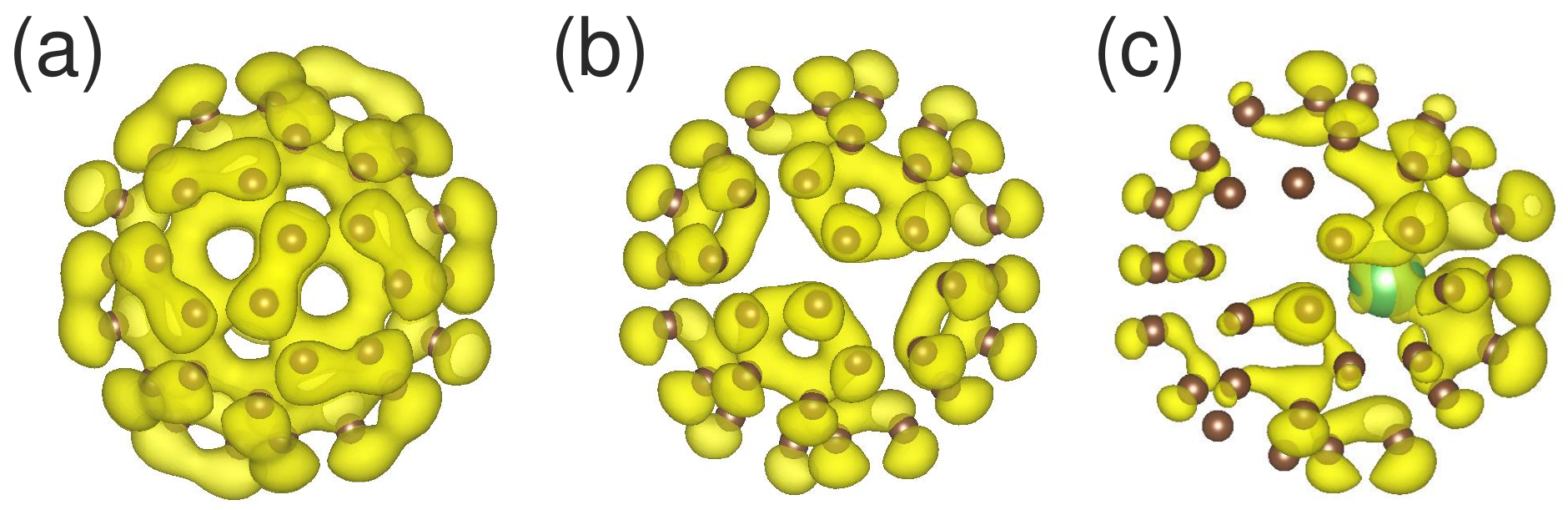}
\caption{%
Partial charge distribution $\rho_{vc}$ of (a) the threefold
degenerate $t_{1u}$ LUMO of C$_{60}$, (b) the fivefold degenerate
$h_u$ HOMO of C$_{60}$, and (c) the partly occupied doubly
degenerate state of La@C$_{60}$ at the Fermi level. The isosurface
value of $\rho_{vc}$ plotted is $0.003$~e/{\AA}$^3$.
% for the $t_{1u}$
%state of C$_{60}$ in (a), $0.005~e/${\AA}$^3$ for the $h_u$ state
%of C$_{60}$ in (b), and $0.002~e/${\AA}$^3$ for the
%doubly-degenerate state of La@C$_{60}$ at the Fermi level in (c).
\label{figS1}}
\end{figure}
%===========< FIGURE S1 >=============================================

%===========< FIGURE S2 >=============================================
% entire page. There is no need to do explicit centering.
\nobreak\quad
\begin{figure*}[t]
\includegraphics[width=1.8\columnwidth]{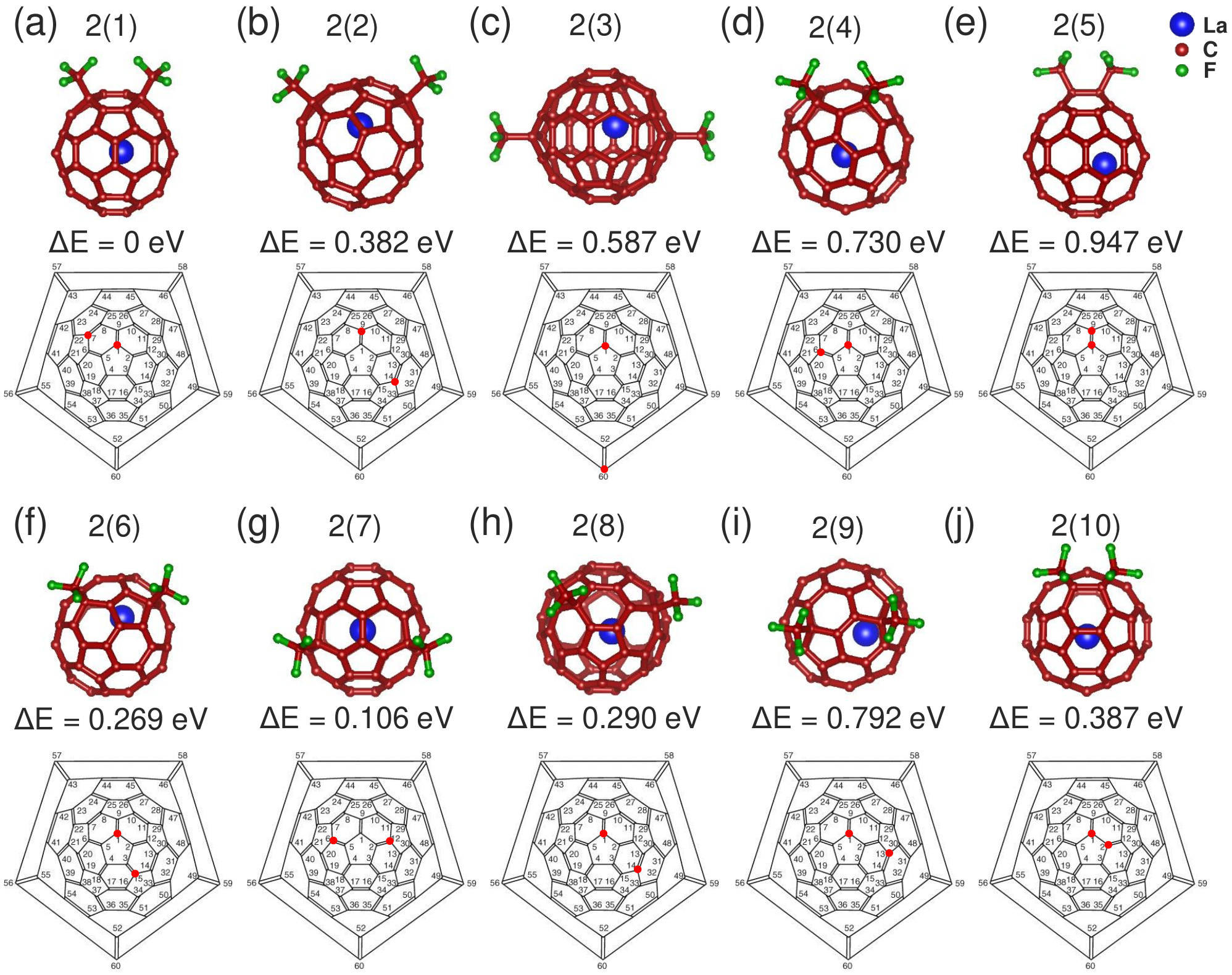}
\caption{Ten different La@C$_{60}$(CF$_3$)$_2$ isomers
functionalized with 2 CF$_3$ radicals. %
The subfigures display the following isomers: %
(a) 2(1), (b) 2(2), (c) 2(3), (d) 2(4), (e) 2(5), (f) 2(6), (g)
2(7), (h) 2(8), (i) 2(9), (j) 2(10). The top panels show the
ball-and-stick models of the structures, followed by the DFT-PBE
relative total energy values ${\Delta}E$ with respect to the most
stable isomer. The bottom panels show the Schlegel diagrams of
functionalized C$_{60}$ molecules, with the
trifluoromethyl sites indicated by the red dots.%
\label{figS2}}
\end{figure*}
%===========< FIGURE S2 >=============================================

%===========< FIGURE S3 >=============================================
% entire page. There is no need to do explicit centering.
\begin{figure*}[t]
\includegraphics[width=1.5\columnwidth]{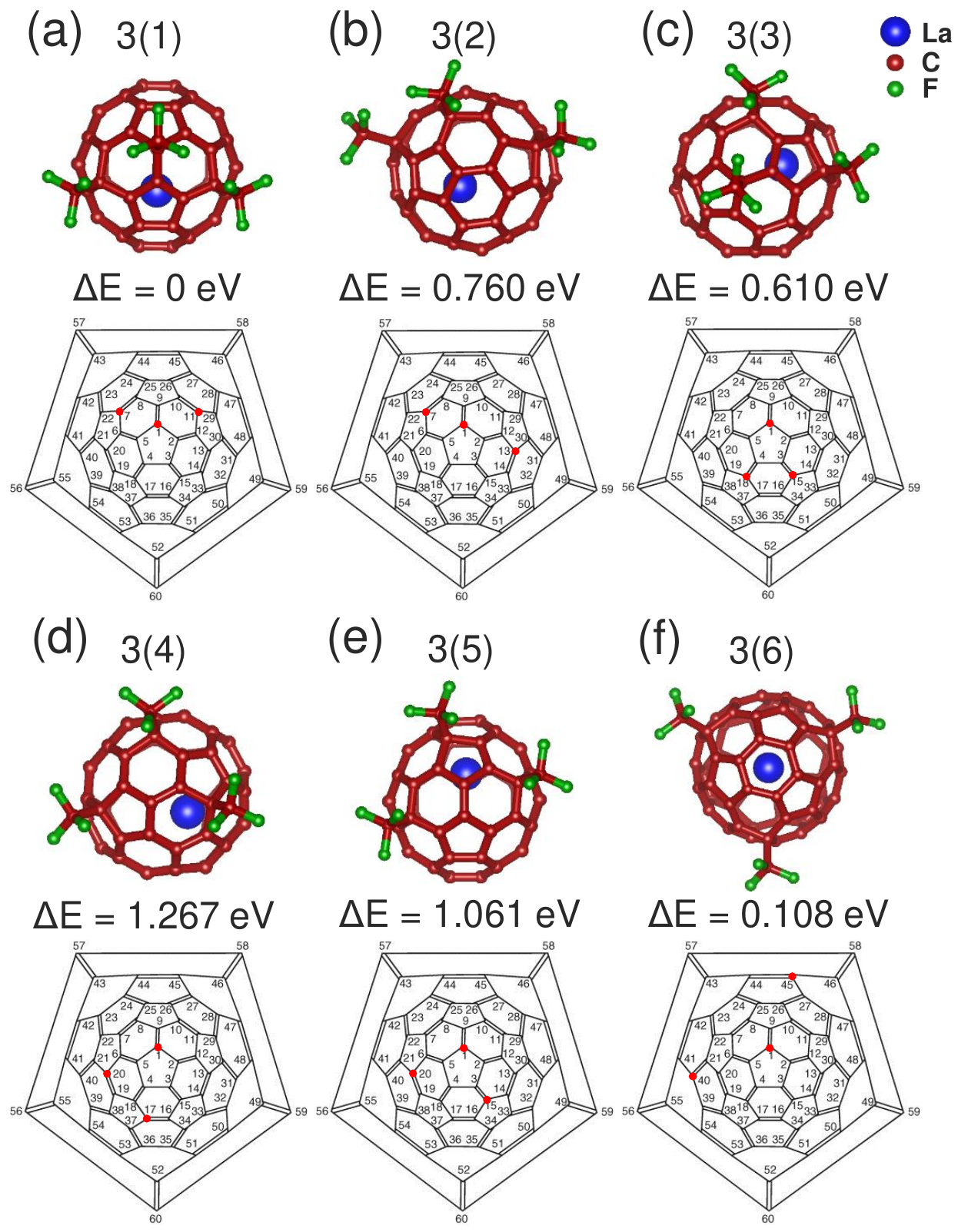}
\caption{Six different La@C$_{60}$(CF$_3$)$_3$ isomers
functionalized with 3 CF$_3$ radicals. %
The subfigures display the following isomers: %
(a) 3(1), (b) 3(2), (c) 3(3), (d) 3(4), (e) 3(5), (f) 3(6). The
top panels show the ball-and-stick models of the structures,
followed by the DFT-PBE relative total energy values ${\Delta}E$
with respect to the most stable isomer. The bottom panels show the
Schlegel diagrams of functionalized C$_{60}$ molecules, with the
trifluoromethyl sites indicated by the red dots.%
\label{figS3}}
\end{figure*}
%===========< FIGURE S3 >=============================================

%===========< FIGURE S4 >=============================================
% entire page. There is no need to do explicit centering.
\begin{figure*}[t]
\includegraphics[width=1.8\columnwidth]{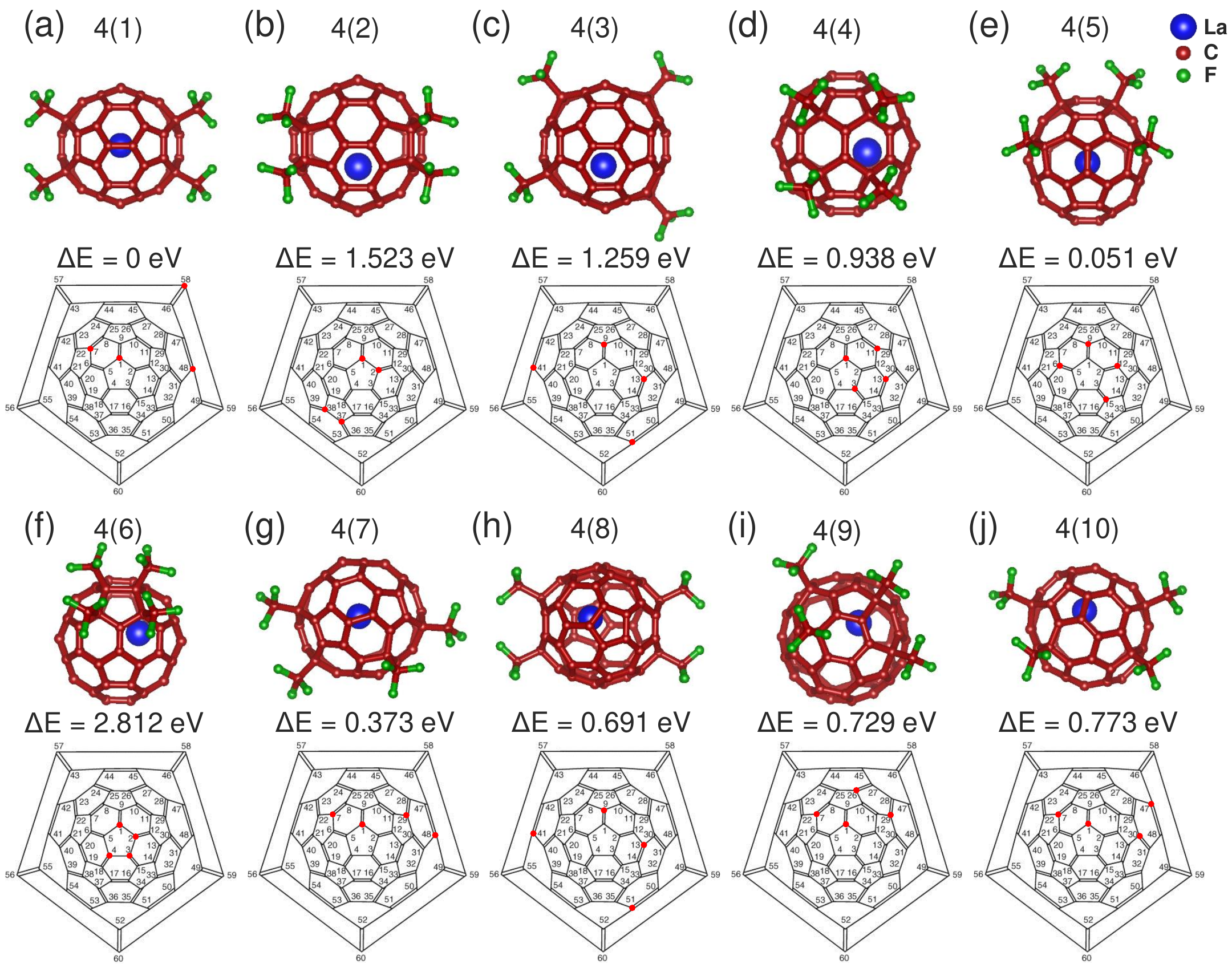}
\caption{Ten different La@C$_{60}$(CF$_3$)$_4$ isomers
functionalized with 4 CF$_3$ radicals. %
The subfigures display the following isomers: %
(a) 4(1), (b) 4(2), (c) 4(3), (d) 4(4), (e) 4(5), (f) 4(6), (g)
4(7), (h) 4(8), (i) 4(9), (j) 4(10). The top panels show the
ball-and-stick models of the structures, followed by the DFT-PBE
relative total energy values ${\Delta}E$ with respect to the most
stable isomer. The bottom panels show the Schlegel diagrams of
functionalized C$_{60}$ molecules, with the
trifluoromethyl sites indicated by the red dots.%
\label{figS4} }
\end{figure*}
%===========< FIGURE S4 >=============================================

%===========< FIGURE S5 >=============================================
% entire page. There is no need to do explicit centering.
\begin{figure*}
\includegraphics[width=1.8\columnwidth]{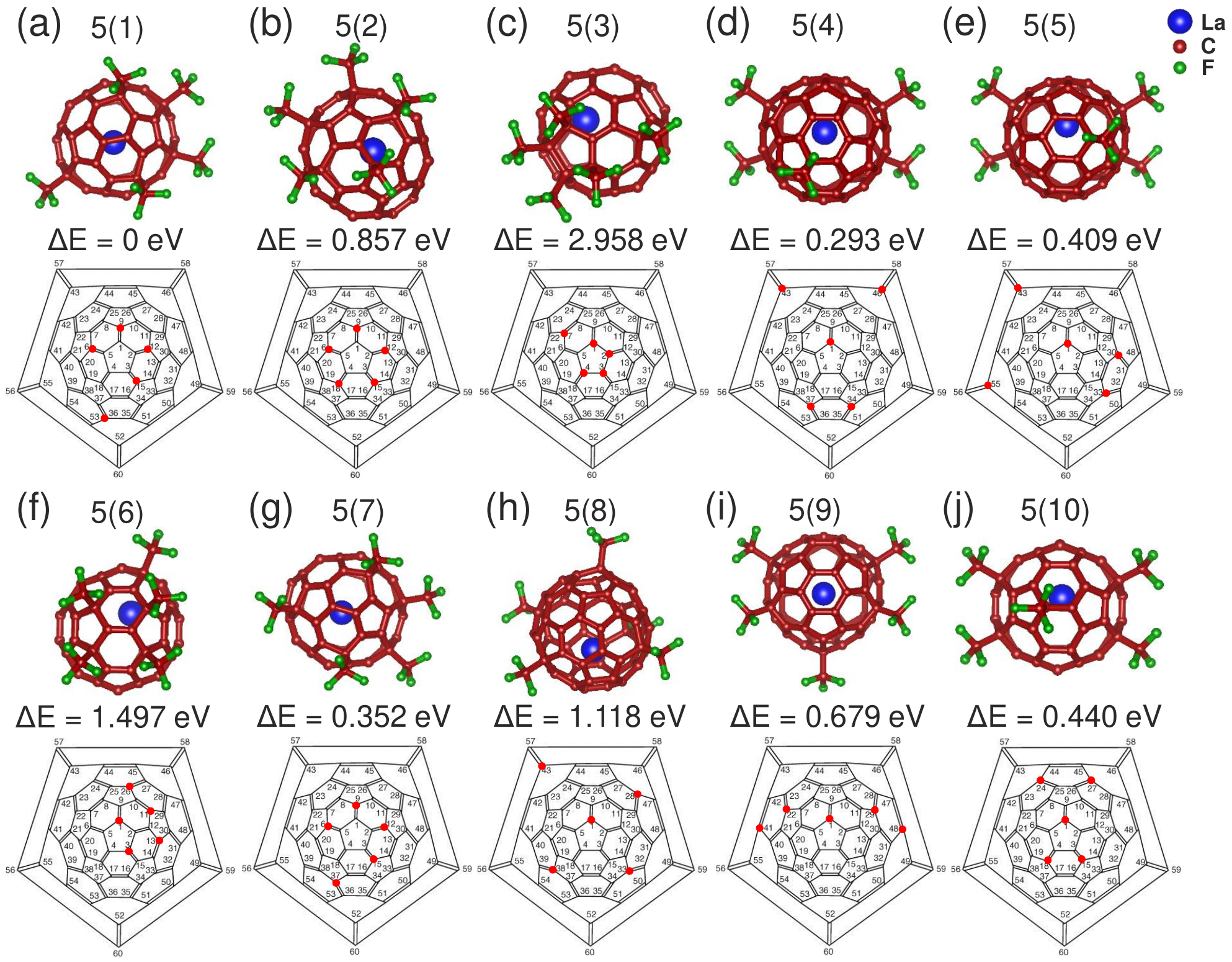}
\caption{Ten different La@C$_{60}$(CF$_3$)$_5$ isomers
functionalized with 5 CF$_3$ radicals. %
(a) 5(1), (b) 5(2), (c) 5(3), (d) 5(4), (e) 5(5), (f) 5(6), (g)
5(7), (h) 5(8), (i) 5(9), (j) 5(10). The top panels show the
ball-and-stick models of the structures, followed by the DFT-PBE
relative total energy values ${\Delta}E$ with respect to the most
stable isomer. The bottom panels show the Schlegel diagrams of
functionalized C$_{60}$ molecules, with the
trifluoromethyl sites indicated by the red dots.%
\label{figS5}}
\end{figure*}
%===========< FIGURE S5 >=============================================

\section{Origin of superconductivity in alkali-based
         M$_3$C$_{60}$ solids}

After a long scientific discussion following the observation of
superconductivity in K$_3$C$_{60}$ and other alkali-based
M$_3$C$_{60}$ intercalation compounds, the following
interpretation of this phenomenon has emerged and is now accepted
by the scientific
community. %
(i) Alkali-based M$_3$C$_{60}$ solids are superconductors
described well by the BCS theory. The electron-phonon coupling
results from a dynamical Jahn-Teller effect on individual $C_{60}$
cages, made possible by retardation, and is confirmed by the
isotope effect observed when substituting pure $^{12}$C$_{60}$ by
pure $^{13}$C$_{60}$ fullerenes. %
(ii) The dominating role of the intercalated alkali atoms is to
partly fill the $t_{1u}$ LUMO of C$_{60}$ that broadens to a
narrow band in the M$_3$C$_{60}$ molecular solid. %
(iii) Changes in $T_c$ are well described by the changing
electron-phonon coupling constant $\lambda=VN(E_F)$ in the
McMillan equation. Since the on-ball Bardeen-Pines interaction $V$
does not change, $\lambda$ is proportional to the electronic
density of states at the Fermi level $N(E_F)$, which is roughly
inversely proportional to the width of the $t_{1u}$-derived band.
Substituting intercalated K atoms by heavier alkali atoms M leads
to an increase of the C$_{60}-$C$_{60}$ separation, thus reducing
the width of the $t_{1u}$ band and consequently increasing the
electron-phonon coupling constant $\lambda$. %
%\hl{%
We should note that the range of lattice constants allowing
superconductivity is limited. Changing the lattice constant
changes the inter-ball hopping integral $t$, while not affecting
the on-ball Coulomb integral $U$. The $U/t$ ratio increases with
increasing lattice constant and, beyond a critical value, changes
doped C$_{60}$ from a metal to a Mott-Hubbard %
insulator {~\cite{{Ganin-10},{Nomura-eC60-15}}}.
%}%

\section{Frontier states in C$_{60}$ and La@C$_{60}$ molecules}

To better understand the effect of the encapsuled La atom on the
electronic structure of the La@C$_{60}$ molecule, we calculated
the partial charge density of the LUMO and the HOMO of C$_{60}$ as
well as that of the doubly degenerate level of La@C$_{60}$ at the
Fermi level. The corresponding results are presented in
Figure{~\ref{figS1}}. As mentioned in the main manuscript, there
is an ${\approx}1.6$~eV wide gap between the threefold degenerate
$t_{1u}$ LUMO and the fivefold degenerate $h_u$ HOMO of the
pristine C$_{60}$ molecule. In La@C$_{60}$, on the other hand, the
partly occupied, doubly degenerate state at $E_F$, which
originally belonged to the threefold degenerate LUMO of C$_{60}$,
defines the Fermi level and provides this molecule with a
``metallic'' character. %

The partial charge densities of the LUMO state in
Figure{~\ref{figS1}}(a) and the HOMO of C$_{60}$ in
Figure{~\ref{figS1}}(b) show that both states are mainly
associated with the p$_\perp$ orbitals of C atoms and are equally
distributed across all atoms of the pristine fullerene molecule.
The partial charge density of the partly occupied state at the
Fermi level of the endohedrally doped La@C$_{60}$ molecule, shown
in Figure{~\ref{figS1}}(c), resembles that of the $t_{1u}$ LUMO of
C$_{60}$. It consists mostly of p$_\perp$ orbitals of C atoms,
with only a small contribution from the enclosed La atom,
consistent with our claim that the main role of La is to transfer
% three
extra electrons to the $\pi$ electron network on the C$_{60}$
cage. We note that the charge in the HOMO/LUMO level of
La@C$_{60}$ is not distributed evenly across the C$_{60}$ cage.
This finding agrees with our Bader charge analysis and is
particularly noticeable in Figure{~\ref{figS1}}(c). It can be
explained by the positively charged La atom gaining energetically
from an off-center displacement, caused by the image-charge
interaction, which skews the electron distribution on the cage
towards the encapsulated La atom. Very similar changes occur in
the $t_{1u}$-derived state of C$_{60}$ that acquires a net charge
from nearby $exohedral$ alkali atoms M in the superconducting
M$_3$C$_{60}$ solid. %

\section{Equilibrium structure and stability of different \\
La@C$_{60}$(CF$_3$)$_m$ isomers}

As mentioned in the main manuscript, the endohedral fullerene
La@C$_{60}$ can be functionalized by CF$_3$ radicals that attach
on-top of C atoms and can be arranged in different ways across the
C$_{60}$ cage. To identify the most stable La@C$_{60}$(CF$_3$)$_m$
geometry, we compared the total energies of different regioisomers
containing $m$ trifluoromethyl radicals. Our results for $m=2,3,4$
and 5 are shown in {Figures}~\ref{figS2}, \ref{figS3}, \ref{figS4}
and \ref{figS5}, respectively. Each isomer is identified as
$m(i)$, where $m$ denotes the number of CF$_3$ radicals and $i$ is
the assigned isomer number.

Our results for 10 different arrangement of $m=2$ CF$_3$ radicals
adsorbed on the C$_{60}$ cage are shown in {Figure}%
~\ref{figS2}. For each regioisomer, we display the optimum
geometry, the corresponding Schlegel diagram and relative energy
with respect to the most stable isomer $i=1$. Our results indicate
that CF$_3$ radicals in the most stable $m=2$ regioisomer are in
the para (third neighbor) positions on a single hexagon on the
C$_{60}$ surface and the molecule has a C$_{2v}$ symmetry. Other
arrangements penalized energetically up to ${\lesssim}1$~eV, with
the least stable arrangement containing CF$_3$ radicals in
adjacent sites. Comparing the relative energies, we found that
CF$_3$ radicals prefer to be close, but not too close on the
C$_{60}$ surface.

As seen in {Figure}~\ref{figS3}, a very similar picture emerges
for $m=3$ CF$_3$ radicals adsorbed on La@C$_{60}$. Comparing the
structure of six different isomers in ball-and-stick models as
well as Schlegel diagrams, we found that the most stable isomer,
shown in {Figure}~\ref{figS3}(a), contains all CF$_3$ radicals in
para (third neighbor) positions on adjacent hexagonal rings on the
C$_{60}$ surface, resulting in a mirror symmetry. The second most
stable isomer, shown in {Figure}~\ref{figS3}(f), contains CF$_3$
radicals separated by 5 neighbor distances, is only
${\approx}0.1$~eV less stable and has a C$_3$ symmetry. Even
though nearest-neighbor arrangements of CF$_3$ radicals were not
considered, other structural candidates incurred an energy penalty
of up to ${\lesssim}1.3$~eV with respect to the most stable
isomer.

The structural paradigm changes for $m=4$ CF$_3$ radicals adsorbed
on the La@C$_{60}$ metallofullerene. Among the 10 regioisomers
displayed in {Figure}~\ref{figS4}, the most stable structure,
shown in {Figure}~\ref{figS4}(a), contains two pairs of CF$_3$
radicals in para-arrangement on hexagonal rings that are separated
by half the circumference of the C$_{60}$ molecule. The
arrangement in {Figure}~\ref{figS4}(e), with all $m=4$ CF$_3$
radicals in para arrangement on adjacent hexagonal rings, is
energetically the second-best isomer, with its energy only
$0.051$~eV higher than the most stable structure. The most stable
isomer has a C$_{2v}$ symmetry and the second most stable isomer
only a mirror symmetry.

The structural paradigm for La@C$_{60}$(CF$_3$)$_m$ regioisomers
with $m=5$ CF$_3$ radicals is similar to the $m=4$ case. Ten
regioisomers are presented in {Figure}~\ref{figS5}. The most
stable of them, shown in {Figure}~\ref{figS5}(a), contains 4
CF$_3$ radicals in para arrangement on three adjacent hexagonal
rings. The last radical is separated by 4 neighbor distances from
the closest CF$_3$ radical. It is also possible to arrange all
five CF$_3$ radicals in para positions on four adjacent hexagonal
rings. As seen in {Figure}~\ref{figS5}(b), this highly symmetric
regioisomer is less stable by ${\lesssim}0.9$~eV than the most
stable structure.

\providecommand{\latin}[1]{#1} \makeatletter \providecommand{\doi}
  {\begingroup\let\do\@makeother\dospecials
  \catcode`\{=1 \catcode`\}=2\doi@aux}
\providecommand{\doi@aux}[1]{\endgroup\texttt{#1}} \makeatother
\providecommand*\mcitethebibliography{\thebibliography} \csname
@ifundefined\endcsname{endmcitethebibliography}
  {\let\endmcitethebibliography\endthebibliography}{}